\documentclass[a4paper,11pt]{article}
\usepackage{jhep-template/jheppub}
\usepackage{amsthm,mathtools,mathrsfs,booktabs,array}
\usepackage{tikz}
\usepackage{float}
\usepackage{needspace}
\usepackage{microtype}
\hypersetup{hypertexnames=false,allcolors=black}
\allowdisplaybreaks
\let\fulltableofcontents\tableofcontents
\renewcommand{\tableofcontents}{{\small\fulltableofcontents}}
\renewcommand{\afterTocRuleSpace}{\clearpage}

\newcommand{\dd}{\mathrm d}
\newcommand{\intk}{\int_{\mathbf k}}
\newcommand{\Ptree}{P_p^{(0)}(t)}
\newcommand{\eval}[1]{\left.#1\right|_{t_2=t_1^-}}
\newcommand{\order}{\mathcal O}
\newcommand{\bluechange}[1]{#1}
\newcommand{\referencechange}[1]{#1}

\definecolor{propagatorred}{RGB}{185,43,47}
\tikzset{
  fprop/.style={draw=propagatorred,line width=0.9pt},
  tvertex/.style={circle,fill=black,draw=black,minimum size=5.4pt,inner sep=0pt},
  svertex/.style={circle,fill=white,draw=black,line width=0.9pt,
    minimum size=5.4pt,inner sep=0pt}
}
\newcommand{\QuarticGraph}[1]{%
  \begin{tikzpicture}[baseline=-0.35cm]
    \useasboundingbox (-1.3,-0.38) rectangle (1.3,0.98);
    \draw[fprop] (-1.2,0)--(0,0)--(1.2,0);
    \draw[fprop] (0,0.39) circle (0.39);
    \node[#1] at (0,0) {};
  \end{tikzpicture}%
}
\newcommand{\FishGraph}[2]{%
  \begin{tikzpicture}[baseline=-0.35cm]
    \useasboundingbox (-1.3,-0.38) rectangle (1.3,0.98);
    \draw[fprop] (-1.2,0)--(-0.55,0);
    \draw[fprop] (0.55,0)--(1.2,0);
    \draw[fprop] (-0.55,0).. controls (-0.22,0.67) and (0.22,0.67)..(0.55,0);
    \draw[fprop] (-0.55,0).. controls (-0.22,-0.67) and (0.22,-0.67)..(0.55,0);
    \node[#2] at (-0.55,0) {};
    \node[#1] at (0.55,0) {};
  \end{tikzpicture}%
}
\newcommand{\TadpoleGraph}[2]{%
  \begin{tikzpicture}[baseline=-0.35cm]
    \useasboundingbox (-1.3,-0.38) rectangle (1.3,0.98);
    \draw[fprop] (-0.57,0)--(0.52,0);
    \draw[fprop] (-0.57,0.39) circle (0.39);
    \draw[fprop] (0.52,0)--(1.2,0.32);
    \draw[fprop] (0.52,0)--(1.2,-0.32);
    \node[#2] at (-0.57,0) {};
    \node[#1] at (0.52,0) {};
  \end{tikzpicture}%
}

\title{Nonlinear gauge maps and the structure of Goldstone loops in canonical inflation}
\author[a,b]{Cheng-Jun Fang}
\author[a,b,c]{and Zong-Kuan Guo}

\affiliation[a]{Institute of Theoretical Physics, Chinese Academy of Sciences,
P.O. Box 2735, Beijing 100190, China}
\affiliation[b]{School of Physical Sciences, University of Chinese Academy of Sciences,
No. 19A Yuquan Road, Beijing 100049, China}
\affiliation[c]{School of Fundamental Physics and Mathematical Sciences,
Hangzhou Institute for Advanced Study, University of Chinese Academy of Sciences,
Hangzhou 310024, China}

\emailAdd{fangchengjun@itp.ac.cn}
\emailAdd{guozk@itp.ac.cn}

\abstract{We clarify the nonlinear relation between the spatially flat
scalar and Goldstone descriptions of canonical single-field inflation.
An exact clock map shows how potential interactions are reorganized
into derivative Goldstone interactions. A change of slicing further
connects their dilation symmetry to its comoving realization. We then
study the Goldstone two-point function at one loop in the decoupling
limit. Using identities derived from the general mode equation, we
reveal how seemingly independent loop contributions combine and their
non-boundary parts cancel once the associated counterterms are
included. This reduction requires no explicit mode-function solution.
The leading adiabatic long-mode response reduces to a total derivative
in loop momentum, which requires an additional regularization
prescription.}

\begin{document}
\maketitle

\section{Introduction}\label{sec:intro}
Transient ultra-slow-roll (USR) evolution can greatly amplify scalar
fluctuations on short scales
\referencechange{\cite{KristianoYokoyama2024Transitions,BarnertEtAl2026}}.
Whether loops involving these modes substantially affect a much longer
adiabatic mode remains under debate
\referencechange{\cite{MotohashiTada2023,FirouzjahiRiotto2024,BallesterosEgea2024}}.
Several explicit calculations report potentially large contributions
\referencechange{\cite{KristianoYokoyama2024,KristianoYokoyama2024Bispectrum,Firouzjahi2023,FrancioliniEtAl2024,Firouzjahi2024Revisiting,ChengLeeNg2024,MaityEtAl2024}},
whereas others argue that the long mode is protected
\referencechange{\cite{Riotto2023,TadaEtAl2024,Inomata2024,KawaguchiEtAl2024}}.
The tension is not only quantitative. The calculations often use
different field variables, in which the interactions and the symmetry
of the long mode take rather different forms.

In the effective field theory (EFT) of inflation, the background
clock is promoted to a Goldstone field $\pi$~\cite{Cheung2008}. Its
time dependence appears directly in the couplings of derivative
interactions. In USR, this formulation has been used for explicit
cubic and quartic loop calculations and for constructing the
nonperturbative Goldstone Hamiltonian
\referencechange{\cite{Firouzjahi2023,ChoudhuryPandaSami2023,FirouzjahiNikbakht2025,FirouzjahiNikbakht2026AllOrders}}.
The same organization of background coefficients supports one-loop
renormalization~\cite{BragliaPinol2025}, studies of superhorizon
freezing~\cite{BragliaPinol2026Freezing}, and calculations for
localized features with hierarchical time dependence
\cite{BragliaCespedesPinol2026}. A relation enforced by dilation
symmetry is less easy to recognize when it is distributed among
individual derivative vertices.

The same problem can instead be organized around the long-wavelength
adiabatic mode. Dilation and soft consistency relations constrain its
effect on short fluctuations
\referencechange{\cite{Maldacena2003,CreminelliZaldarriaga2004,HinterbichlerEtAl2012,GoldbergerHuiNicolis2013,HinterbichlerEtAl2014}}.
Loop-level conservation has been argued for on this basis in
single-clock settings
\cite{PimentelEtAl2012,AssassiEtAl2013,SenatoreZaldarriaga2013AllLoops}.
For transient non-slow-roll backgrounds, soft-theorem, path-integral,
and \referencechange{nonlinear} analyses have likewise argued for cancellations
\referencechange{\cite{TadaEtAl2024,Inomata2024,KawaguchiEtAl2024,Fumagalli2025PartI,Fumagalli2025,ArtigasEtAl2026}}.
Here the long-mode response supplies the organizing principle, rather
than any individual loop vertex.

The proposed cancellations also take different forms in different
representations. In comoving gauge, Ref.~\cite{Fumagalli2025}
expresses the combined cubic and quartic contribution as a total
derivative in loop momentum. Flat-gauge analyses find related Ward
and mode-equation constraints~\cite{FangLyuChenGuo2026,FangGuo2026}.
The homogeneous response introduces another set of terms: the role of
backreaction, tadpoles, and their counterterms has been examined in
Refs.~\cite{Inomata2025Backreaction,FangEtAl2025Backreaction,Inomata2025Counterterms,KristianoYokoyama2025Background}.
Separate-universe arguments and a soft-curvature EFT provide further
ways of describing the long-short relation
\referencechange{\cite{IacconiEtAl2024,IacconiEtAl2026,EmaEtAl2026,TanakaUrakawa2026}}.

\referencechange{What is less transparent is how these descriptions are connected
in the derivative Goldstone theory.} How does the flat-gauge scalar become the EFT
Goldstone? Where does its compensating time shift go when the slicing
is changed? And can the associated dilation structure be seen
directly in the Goldstone loop calculation?

The starting point is an exact nonlinear clock map on a fixed FRW
background. In the flat scalar variable $\delta\phi$, nonlinear
interactions are potential-like; expressed through the background
clock, the same dynamics acquires the derivative interactions of the
Eulerian Goldstone $\pi$. The flat-slicing dilation shifts both time
and the fluctuation. Taking the invariant clock as the time coordinate
absorbs the compensating time shift into the slicing, and the
long-wavelength curvature perturbation $\zeta$ acquires the familiar
comoving dilation. The apparently different fields and symmetry
transformations are thus connected by explicit nonlinear maps.

\referencechange{This relation guides our one-loop calculation, which reveals
how apparently independent Goldstone derivative-vertex diagrams
are related.} The cubic graphs alone do not display the final
structure: integration over the earlier vertex separates an
equal-time endpoint from a history term. The endpoint is matched by
the quartic Hamiltonian, including its Legendre contact. A dilation
identity derived from the free mode equation brings the nonlocal
history term into the same equal-time language as the local
contributions, without requiring a particular mode-function solution.
The remaining local response is completed by the tadpole sector and
the counterterms fixed through the one-point condition.
\referencechange{Together these terms reduce the leading adiabatic response
to a total derivative in loop momentum, providing a Goldstone
counterpart to structures obtained in other field
descriptions~\cite{Fumagalli2025,FangGuo2026,EmaEtAl2026}.}
We leave this total derivative
unevaluated. This separates the cancellation of non-boundary terms
from the prescription needed to assign a value to the momentum
integral.

The paper is organized as follows. Section~\ref{sec:eft} sets up the
canonical EFT and decoupling limit, while section~\ref{sec:gauge}
develops the clock and slicing maps. The Goldstone modes, vertices,
and diagrams are organized in section~\ref{sec:scope}; their
long-wavelength reduction is carried out in
section~\ref{sec:cancellation}. Section~\ref{sec:discussion} discusses
the result, with technical derivations collected in the appendices.

\section{Inflationary EFT and the decoupling limit}\label{sec:eft}
The homogeneous inflaton selects a preferred time slicing. In unitary
gauge the leading operators of the EFT of inflation can be written as
\begin{align}\label{eq:eft-unitary}
S=\int\dd^4x\sqrt{-g}\bigg[&
\frac{M_{\rm Pl}^2}{2}R+M_{\rm Pl}^2\dot H(t)g^{00}
-M_{\rm Pl}^2\bigl(3H^2(t)+\dot H(t)\bigr)
\nonumber\\
&+\sum_{n\geq2}\frac{M_n^4(t)}{n!}(\delta g^{00})^n+\cdots
\bigg],
\end{align}
where $\delta g^{00}\equiv g^{00}+1$ and the ellipsis denotes further operators allowed by spatial
diffeomorphisms~\cite{Cheung2008}. We restrict to the canonical
single-field sector, with $M_n=0$ and no independent higher-derivative
operators. We take a smooth inflationary background, whose equations are
\begin{equation}\label{eq:bg-eqs}
3M_{\rm Pl}^2H^2=\frac12\Pi^2+V(\bar\phi),\qquad
-M_{\rm Pl}^2\dot H=\frac12\Pi^2,\qquad
\dot\Pi+3H\Pi+V^{(1)}(\bar\phi)=0,
\end{equation}
with $\Pi\equiv\dot{\bar\phi}$. Restoring the broken time
diffeomorphism amounts to replacing $t$ by $t+\pi$ in the coefficients
and $g^{00}$ by
$g^{\mu\nu}\partial_\mu(t+\pi)\partial_\nu(t+\pi)$.
This gives the Goldstone description used below
\cite{Cheung2008,BragliaPinol2025,BragliaCespedesPinol2026}.

We work at leading order in the scalar-gravity decoupling expansion,
keeping the background FRW metric fixed while retaining the leading
nonzero scalar interactions. Equivalently, in the present canonical
sector we keep the terms linear in
$A\equiv-M_{\rm Pl}^2\dot H$ and drop terms with an additional
explicit factor of $\epsilon=-\dot H/H^2$. This does not require
$\eta\equiv\dot\epsilon/(H\epsilon)$ to be small:
$\dot A/A=H(\eta-2\epsilon)$ may be of order $H$, and its time
derivatives are kept. The counting assumes that higher derivatives of
$H$ do not compensate the explicit extra power of $\epsilon$.
This accommodates a transient ultra-slow-roll phase and fixes the
accuracy of all symmetry checks and the loop calculation below.
The same decoupling organization underlies recent EFT loop
calculations~\cite{BragliaPinol2025,BragliaCespedesPinol2026}.

\section{Nonlinear clock map and gauge equivalence}\label{sec:gauge}
\subsection{From the flat scalar to the Eulerian Goldstone}
In spatially flat slicing, $\phi=\bar\phi+\delta\phi$ on the fixed
FRW geometry. The scalar matter action is
\begin{equation}\label{eq:flat-exact}
S_{\rm flat}=\int\dd t\,\dd^3x\,a^3
\left[\frac12(\Pi+\delta\dot\phi)^2
-\frac{(\partial_i\delta\phi)^2}{2a^2}
-V(\bar\phi+\delta\phi)\right].
\end{equation}
Using Eq.~\eqref{eq:bg-eqs}, its fluctuation bulk has the simple form
\begin{equation}\label{eq:flat-fluct}
S_{\rm flat,fluct}=\int\dd t\,\dd^3x\,a^3
\left[\frac12\delta\dot\phi^2
-\frac{(\partial_i\delta\phi)^2}{2a^2}
-\sum_{n\geq2}\frac{V^{(n)}(\bar\phi)}{n!}\delta\phi^n\right],
\end{equation}
after removing the background and the integral of
$\dd(a^3\Pi\delta\phi)/\dd t$. In this variable the nonlinear
vertices are potential vertices, whereas the Goldstone action below
contains time and spatial derivatives at every order.
On an interval where $\Pi\ne0$, the background clock is invertible
and defines the Eulerian Goldstone nonlinearly by
\begin{equation}\label{eq:clock-map}
\phi(t,\mathbf x)=\bar\phi(t+\pi(t,\mathbf x)),\qquad
\delta\phi=\Pi\pi+\frac12\dot\Pi\pi^2
+\frac16\ddot\Pi\pi^3+\cdots .
\end{equation}
The label $\pi$ in the rest of this paper denotes the flat/Eulerian
field $\pi_{\rm E}$ of the clock map. Direct substitution into
Eq.~\eqref{eq:flat-exact}, without using a background equation, gives
\begin{equation}\label{eq:eulerian-exact}
S_{\rm E}=\int\dd t\,\dd^3x\,a^3\left\{
\frac12\Pi^2(t+\pi)
\left[(1+\dot\pi)^2-\frac{(\partial_i\pi)^2}{a^2}\right]
-V\!\left(\bar\phi(t+\pi)\right)\right\}.
\end{equation}
Thus the two fixed-background actions are related by an exact local
field redefinition.

The Friedmann equations turn Eq.~\eqref{eq:eulerian-exact} into a
particularly useful bulk-plus-boundary decomposition. The identity
$(1+\dot\pi)\dot H(t+\pi)=\dd H(t+\pi)/\dd t$ yields
\begin{align}\label{eq:eulerian-ibp}
S_{\rm E}={}&3M_{\rm Pl}^2\int\dd t\,\dd^3x\,a^3H^2(t)
-2M_{\rm Pl}^2\int\dd t\,\dd^3x\,
\frac{\dd}{\dd t}\bigl[a^3H(t+\pi)\bigr]
\nonumber\\
&+\int\dd t\,\dd^3x\,a^3\left\{
A(t+\pi)\left[\dot\pi^2-
\frac{(\partial_i\pi)^2}{a^2}\right]
-3M_{\rm Pl}^2\bigl[H(t+\pi)-H(t)\bigr]^2
\right\}.
\end{align}
The first term is background only. The second is the boundary term
generated by integrating the Goldstone Lagrangian by parts. For its
treatment in inflationary correlators, see
Refs.~\cite{Fumagalli2025PartI,BragliaPinol2024,ArrojaTanaka2011,WangEtAl2025FieldRedefinitions,MaEtAl2026Boundary}.
The last, non-derivative
term starts at $-3M_{\rm Pl}^2\dot H^2\pi^2$ and is suppressed by an
additional power of $\epsilon$ in the counting of
section~\ref{sec:eft}.
Hence the leading bulk action takes the all-order
form~\cite{Creminelli:2024cge}
\begin{equation}\label{eq:leading-bulk}
S_{\rm dec}=\int\dd t\,\dd^3x\,a^3 A(t+\pi)
\left[\dot\pi^2-\frac{(\partial_i\pi)^2}{a^2}\right].
\end{equation}
In the flat variable, the nonlinearity appears in the
potential of Eq.~\eqref{eq:flat-fluct}. The clock redefinition
represents the same fixed-background dynamics through derivative
interactions generated by $A(t+\pi)$. The cancellations below reflect
the underlying equivalence of the two descriptions.
Expanding $A(t+\pi)$ to second order in $\pi$ gives all the cubic
and quartic vertices used in section~\ref{sec:scope}.

\subsection{From flat slicing to the comoving dilation}\label{sec:slicing-dilation}
The replacement of $\delta\phi$ by $\pi$ in
Eq.~\eqref{eq:clock-map} changes the field variable, not the
coordinates: $\pi(t,\mathbf x)$ is still defined on spatially
flat slices. In particular, the argument $t+\pi$ has not yet
been adopted as a time coordinate. For an infinitesimal constant
$\lambda$, the flat-slicing action has the symmetry
\cite{FangGuo2026}
\begin{equation}\label{eq:dilation-map}
\widetilde{\mathbf x}=(1-\lambda)\mathbf x,\qquad
\widetilde t=t+\frac{\lambda}{H(t)},\qquad
\delta\widetilde\phi(\widetilde t,\widetilde{\mathbf x})
=\delta\phi(t,\mathbf x)-\lambda\frac{\Pi(t)}{H(t)}.
\end{equation}
The clock map induces the Goldstone transformation
\begin{equation}\label{eq:dilation-clocks}
\widetilde\pi(\widetilde t,\widetilde{\mathbf x})
=\pi(t,\mathbf x)-\frac\lambda{H(t)}.
\end{equation}
Indeed, the total scalar and its clock argument obey
\begin{equation}\label{eq:flat-clock-invariants}
\bar\phi(\widetilde t)+\delta\widetilde\phi
\simeq\bar\phi(t)+\delta\phi,\qquad
\widetilde t+\widetilde\pi=t+\pi.
\end{equation}
The scale factor satisfies $a(\widetilde t)\simeq(1+\lambda)a(t)$,
while $\dd(\lambda/H)/\dd t=\order(\lambda\epsilon)$. Hence
\begin{equation}\label{eq:flat-dilation-factors}
a^3(t)\,\dd t\,\dd^3x
\simeq a^3(\widetilde t)\,\dd\widetilde t\,
\dd^3\widetilde x,\qquad
\frac{\partial_i}{a(t)}
\simeq\frac{\widetilde\partial_i}{a(\widetilde t)}.
\end{equation}
These relations preserve the kinetic and potential terms of
Eqs.~\eqref{eq:flat-exact} and \eqref{eq:eulerian-exact}, giving
\begin{equation}\label{eq:flat-actions-symmetry}
S_{\rm flat}[\delta\phi]\simeq S_{\rm flat}[\delta\widetilde\phi],
\qquad
S_{\rm E}[\pi]\simeq S_{\rm E}[\widetilde\pi],
\end{equation}
with the integration domains transformed along with the coordinates.

In comoving gauge, residual spatial dilation instead shifts the
curvature perturbation $\zeta$ by a constant at fixed comoving
time~\cite{HinterbichlerEtAl2012}. Its relation to the cancellation
of large-scale loop corrections has been studied, for example, in
Refs.~\cite{Fumagalli2025,EmaEtAl2026}. The flat-slicing
transformations above include a compensating time shift that is
absent from this comoving representation. To relate the two
symmetries, we must therefore change the slicing, not merely
rename the field.

{\color{black}
We now use the invariant clock argument as the time coordinate $t_1$.
Where this change is locally invertible, define its inverse
displacement $\pi_1$ by
\begin{equation}\label{eq:inverse-clock}
t_1=t+\pi(t,\mathbf x),\qquad
t=t_1-\pi_1(t_1,\mathbf x),\qquad
\pi_1(t_1,\mathbf x)=\pi(t,\mathbf x).
\end{equation}
The scalar then satisfies $\phi(t_1,\mathbf x)=\bar\phi(t_1)$:
constant-$t_1$ surfaces are uniform-inflaton, or comoving, slices.
The scalar fluctuation has not disappeared. It is now
encoded in the displacement between the comoving clock surface and
the original flat slice, measured by $\pi_1$. The spatial coordinates
have not yet been changed.
\par}

\Needspace{9\baselineskip}
\bluechange{In these variables the exact fixed-background action takes
the form}
\begin{align}\label{eq:inverse-action}
S[\pi_1]=\int\dd t_1\,\dd^3x\,a^3(t_1-\pi_1)
\bigg\{&\frac{\Pi^2(t_1)}{2(1-\dot{\pi}_1)}
\left[1-\frac{(\partial_i\pi_1)^2}{a^2(t_1-\pi_1)}\right]
\nonumber\\
&-(1-\dot{\pi}_1)V\!\left(\bar\phi(t_1)\right)\bigg\}.
\end{align}
\bluechange{Its derivation from the inverse-clock Jacobian and
derivative map is given in App.~\ref{app:inverse-clock-map}.}

\bluechange{Because $t_1$ is the invariant clock argument, the same
dilation now acts at fixed $t_1$:}
\begin{equation}\label{eq:inverse-clock-dilation}
\widetilde{\mathbf x}=(1-\lambda)\mathbf x,\qquad
\tilde{\pi}_1(t_1,\widetilde{\mathbf x})
\simeq\pi_1(t_1,\mathbf x)-\frac{\lambda}{H(t_1)}.
\end{equation}
\bluechange{The compensating time shift of the flat-slicing
representation has been absorbed into the clock coordinate. At the
retained order the action satisfies $S[\tilde\pi_1]\simeq S[\pi_1]$.
The measure and derivative check is given in
App.~\ref{app:inverse-clock-map}.}

The connection to $\zeta$ follows from the spatial geometry.
On a constant-$t_1$ slice, the original FRW metric induces
\begin{equation}\label{eq:inverse-induced-metric}
h_{ij}=a^2(t_1-\pi_1)\delta_{ij}
-\partial_i\pi_1\partial_j\pi_1.
\end{equation}
The variable $\pi_1$ is therefore a comoving-clock variable, not
the full comoving-gauge curvature perturbation: a spatial
threading transformation is still needed to put the metric in
the standard comoving-gauge form
\cite{Maldacena2003,BehbahaniEtAl2012}. For the long-wavelength
geometry, neglecting spatial-gradient corrections in this
identification gives~\cite{BragliaPinol2025,BehbahaniEtAl2012}
\begin{equation}\label{eq:inverse-zeta-map}
\zeta(t_1,\mathbf x)
\simeq\ln\frac{a(t_1-\pi_1)}{a(t_1)}
=-\int_{t_1-\pi_1}^{t_1}H(s)\,\dd s
\simeq-H(t_1)\pi_1(t_1,\mathbf x).
\end{equation}
At linear order, $\pi_1=\pi$, reproducing the standard EFT
relation $\zeta=-H\pi$~\cite{Cheung2008}.
Equation~\eqref{eq:inverse-clock-dilation} then implies
\begin{equation}\label{eq:comoving-dilation}
\tilde\zeta(t_1,\widetilde{\mathbf x})
\simeq-H(t_1)\tilde{\pi}_1(t_1,\widetilde{\mathbf x})
\simeq\zeta(t_1,\mathbf x)+\lambda.
\end{equation}
\bluechange{This is the standard comoving dilation. The flat-slicing
and comoving transformations are the same large symmetry expressed
on different slices.}

\section{Goldstone modes and one-loop organization}\label{sec:scope}
This section organizes the one-loop Goldstone two-point
function at finite external momentum. We first specify the free
modes and interaction vertices. The one-loop topologies are then
written with their derivative structure explicit. This prepares
the diagram-by-diagram soft-limit analysis of
Sec.~\ref{sec:cancellation}.

\subsection{Free theory and interactions}\label{sec:free}
We use the Eulerian Goldstone $\pi$ on a fixed FRW background and define
\begin{equation}\label{eq:defs}
A=-M_{\rm Pl}^2\dot H>0,\qquad C=a^3A,\qquad
\alpha=\frac{\dot A}{A},\qquad \omega_k^2=\frac{k^2}{a^2}.
\end{equation}
Our momentum-integral notation is
$\intk\equiv\int\dd^3k/(2\pi)^3$.
The derivative bulk action through fourth order is
\begin{equation}\label{eq:action}
S_{\rm bulk}=\int\dd t\,\dd^3x\,
\left[a^3K(\pi,t)\dot\pi^2-aK(\pi,t)(\partial_i\pi)^2\right],
\qquad K=A+\dot A\pi+\tfrac12\ddot A\pi^2.
\end{equation}
The free modes are specified by
\begin{equation}\label{eq:mode}
\partial_t(C\dot u_k)+C\omega_k^2u_k=0,\qquad
2C(u_k\dot u_k^*-u_k^*\dot u_k)=i.
\end{equation}
We write $W_k(t_1,t_2)=u_k(t_1)u_k^*(t_2)$,
$F_k=\mathrm{Re}\,W_k$, and
$R_k(t_1,t_2)=2\theta(t_1-t_2)\mathrm{Im}\,W_k$.
At a vertex time $t$, the equal-time mode contractions are as
follows; we suppress their arguments below when the vertex time is clear:
\begin{equation}\label{eq:Ldefs}
L_{00}(k,t)=|u_k(t)|^2,\qquad
L_{01}(k,t)=\mathrm{Re}[u_k(t)\dot u_k^*(t)],\qquad
L_{11}(k,t)=|\dot u_k(t)|^2.
\end{equation}
The prime on a two-point function removes
$(2\pi)^3\delta^{(3)}(\mathbf p+\mathbf p')$:
$\Ptree=\langle\pi_{\mathbf p}(t)\pi_{-\mathbf p}(t)\rangle_0'
=|u_p(t)|^2$.

The bulk interaction Hamiltonian contains the two cubic and two quartic
vertices~\cite{FirouzjahiNikbakht2025}
\begin{align}\label{eq:vertices}
H_{3,T}&=g_T\!\int\dd^3x\,\pi\dot\pi^2,&
H_{3,S}&=g_S\!\int\dd^3x\,\pi(\partial_i\pi)^2,\\
H_{4,K}&=\lambda_K\!\int\dd^3x\,\pi^2\dot\pi^2,&
H_{4,S}&=\lambda_S\!\int\dd^3x\,\pi^2(\partial_i\pi)^2,
\end{align}
where every field in a vertex is an interaction-picture field and
\begin{equation}\label{eq:coeffs}
g_T=-C\alpha,\quad g_S=\frac{C\alpha}{a^2},\qquad
\lambda_K=\frac C2(\alpha^2-\dot\alpha),\quad
\lambda_S=\frac{C}{2a^2}(\alpha^2+\dot\alpha).
\end{equation}
The kinetic quartic coefficient includes the contact generated by the
Legendre transform. Its derivation is given in App.~\ref{app:ham}.

For the in-in integrals, we choose the Bunch--Davies state in an
initial scaling phase with constant $A$, while $H$ may be treated as
constant at the retained decoupling order. Taking $t_i$
in that phase gives $\alpha_i=\dot\alpha_i=0$; the initial surface
is treated with the associated adiabatic contour prescription.

\subsection{One-loop integrals and Feynman diagrams}\label{sec:bare}
The bulk in-in expansion of Eq.~\eqref{eq:vertices} follows the
standard organization of inflationary loop correlators
\cite{Weinberg2005,Weinberg2006,SenatoreZaldarriaga2010} and gives
\begin{equation}\label{eq:topologies}
\delta P_{\rm bare}=\delta P_{4K}+\delta P_{4S}
+\sum_{X,Y\in\{T,S\}}\delta P_{XY}^{\rm fish}
+\sum_{X,Y\in\{T,S\}}\delta P_{XY}^{\rm tad/1PR}.
\end{equation}
Here $X$ labels the later cubic vertex and $Y$ the earlier one; $TS$
and $ST$ are distinct ordered graphs. The tadpole terms are connected
contractions of the two-point function in which a zero-momentum bridge
joins an external-leg vertex to a self-contracted vertex. We give
the finite-$p$ results below. Their derivation, auxiliary kernels,
and derivative conventions are collected in App.~\ref{app:wick}.

The connected kernels enter the bare correlator as
\begin{align}\label{eq:kernel-integrals}
\delta P_{4X}
&=\int_{t_i}^{t}\dd t_1\intk\mathcal I_{4,X}(p,k;t,t_1),\\
\delta P_{XY}^{\rm fish}
&=\int_{t_i}^{t}\dd t_1\int_{t_i}^{t_1}\dd t_2\intk
\mathcal I_{XY}(p,k;t,t_1,t_2),\\
\delta P_{XY}^{\rm tad/1PR}
&=\int_{t_i}^{t}\dd t_1\int_{t_i}^{t_1}\dd t_2\intk
\mathcal I_{XY}^{\rm tad/1PR}(p,k;t,t_1,t_2).
\end{align}
The first line has $X\in\{K,S\}$, while the two-vertex terms have
$X,Y\in\{T,S\}$.

For a quartic vertex at $t_1$, let
$E_0=W_p(t,t_1)$, $E_1=\partial_{t_1}W_p(t,t_1)$, and
$L_{ab}=L_{ab}(k;t_1)$.  The two quartic kernels are
\begin{align}\label{eq:quartickernels}
\mathcal I_{4,K}&=2\mathrm{Im}\!\left\{\lambda_K
\left(2E_1^2L_{00}+2E_0^2L_{11}+8E_0E_1L_{01}\right)\right\},\\
\mathcal I_{4,S}&=2\mathrm{Im}\!\left\{
2\lambda_S(p^2+k^2)E_0^2L_{00}\right\}.
\end{align}
The derivative placements that distinguish the one-loop
terms remain explicit in the kernels. The diagrams below record only
their topology and vertex type. This separation matters in the soft
limit, where the free mode equation recombines the time- and
spatial-derivative terms.
\begin{figure}[htbp]
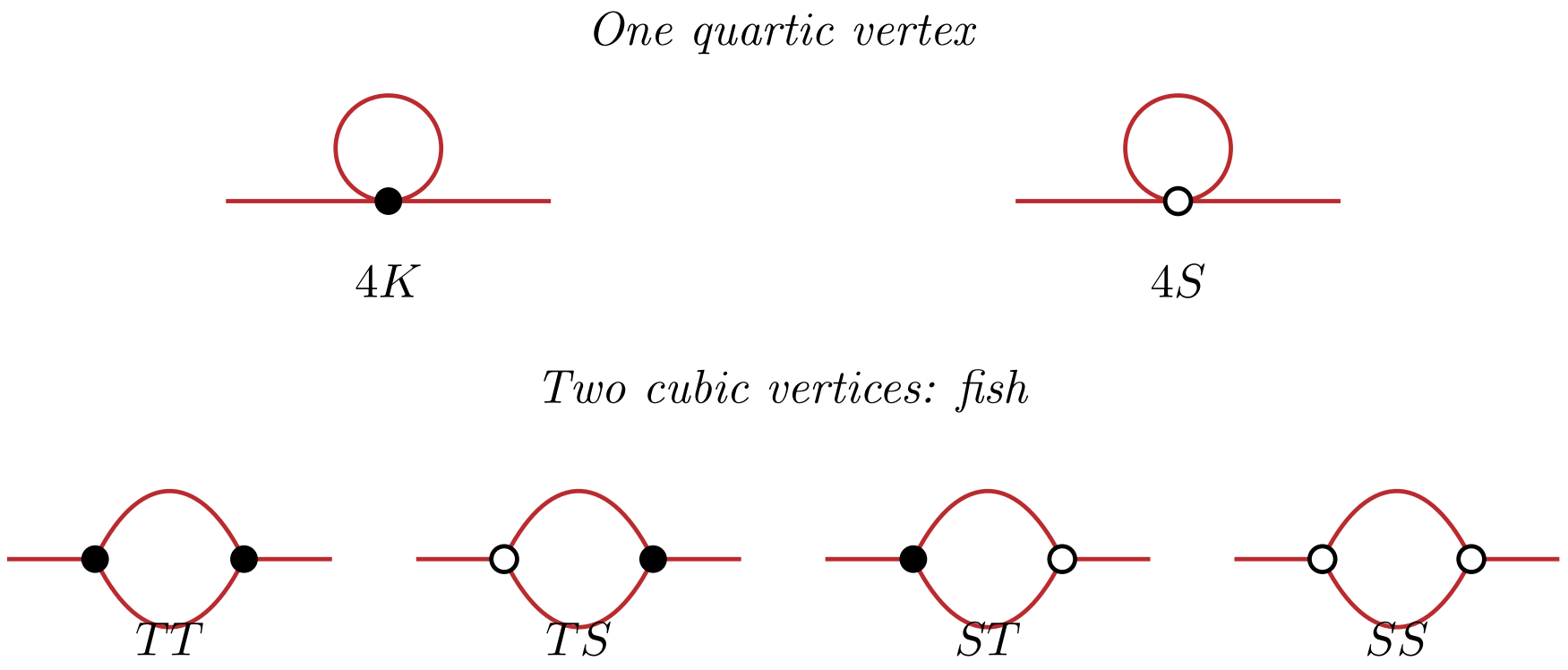

\centering
\small
\begin{tabular}{@{}cccc@{}}
\multicolumn{4}{c}{\emph{One quartic vertex}}\\[1pt]
\multicolumn{2}{c}{\QuarticGraph{tvertex}}&
\multicolumn{2}{c}{\QuarticGraph{svertex}}\\[-3pt]
\multicolumn{2}{c}{$4K$}&\multicolumn{2}{c}{$4S$}\\[9pt]
\multicolumn{4}{c}{\emph{Two cubic vertices: fish}}\\[1pt]
\FishGraph{tvertex}{tvertex}&\FishGraph{tvertex}{svertex}&
\FishGraph{svertex}{tvertex}&\FishGraph{svertex}{svertex}\\[-3pt]
$TT$&$TS$&$ST$&$SS$\\[9pt]
\end{tabular}
\caption{The quartic and fish classes in Eq.~\eqref{eq:topologies}. Red lines
are free $\pi$ propagators; black and white dots denote time- and
spatial-derivative vertices, respectively. In two-vertex graphs the left
dot is the earlier vertex $Y$ at $t_2$, and the right dot is the later
vertex $X$ at $t_1$, so the label is $XY$. The quartic $4K$ vertex
includes its Legendre contact. The sketches specify topology and vertex
type; the Wick kernels specify which incident lines are differentiated.}
\label{fig:one-loop-graphs}
\end{figure}

The four ordered fish kernels use
$\mathbf q=\mathbf p-\mathbf k$ and
$\mathcal K_{pkq}=p^2+k^2+q^2$. The propagator products
are
\begin{align}\label{eq:fishbase}
\mathscr K_{Q12}&=W_p(t,t_1)W_p(t,t_2)
W_k(t_1,t_2)W_q(t_1,t_2),\\
\mathscr K_{2Q1}&=W_p(t,t_1)W_p^*(t,t_2)
W_k^*(t_1,t_2)W_q^*(t_1,t_2).
\end{align}
For the external line $e$ and internal lines $k,q$ at vertex $j$,
define
\begin{equation}\label{eq:Dt}
\mathbb D_j=\partial_{j,e}\partial_{j,k}
+\partial_{j,e}\partial_{j,q}
+\partial_{j,k}\partial_{j,q}.
\end{equation}
Each $\partial_{j,\ell}$ differentiates the Wightman factor on
line $\ell$ with respect to $t_j$, holding the vertex coefficient
fixed. The four kernels are
\begin{align}\label{eq:fishkernels}
\mathcal I_{TT}&=2\mathrm{Re}\left\{
4g_{T1}g_{T2}\mathbb D_1\mathbb D_2
(\mathscr K_{2Q1}-\mathscr K_{Q12})\right\},\\
\mathcal I_{TS}&=2\mathrm{Re}\left\{
2g_{T1}g_{S2}\mathcal K_{pkq}\mathbb D_1
(\mathscr K_{2Q1}-\mathscr K_{Q12})\right\},\\
\mathcal I_{ST}&=2\mathrm{Re}\left\{
2g_{S1}g_{T2}\mathcal K_{pkq}\mathbb D_2
(\mathscr K_{2Q1}-\mathscr K_{Q12})\right\},\\
\mathcal I_{SS}&=2\mathrm{Re}\left\{
g_{S1}g_{S2}\mathcal K_{pkq}^2
(\mathscr K_{2Q1}-\mathscr K_{Q12})\right\}.
\end{align}
\begin{figure}[htbp]
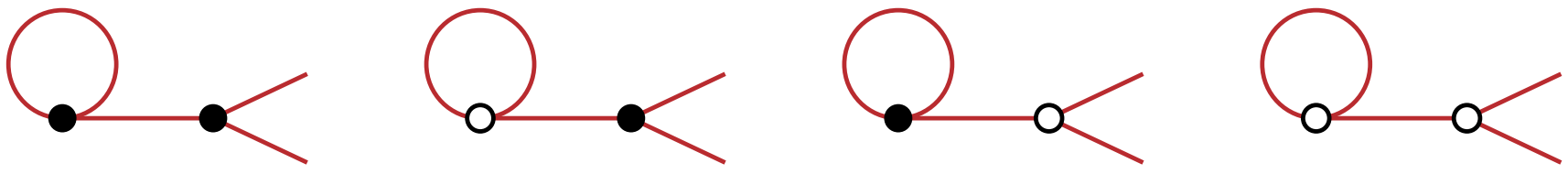

\centering
\small
\begin{tabular}{@{}cccc@{}}
\TadpoleGraph{tvertex}{tvertex}&\TadpoleGraph{tvertex}{svertex}&
\TadpoleGraph{svertex}{tvertex}&\TadpoleGraph{svertex}{svertex}\\[-3pt]
$TT$&$TS$&$ST$&$SS$
\end{tabular}
\caption{The four connected non-1PI tadpole classes in
Eq.~\eqref{eq:topologies}, using the convention of
Fig.~\ref{fig:one-loop-graphs}. The earlier vertex $Y$ on the left is
self-contracted; its zero-momentum bridge reaches the later vertex $X$,
which carries both external legs. Each label lists the later vertex
first. These graphs are direct connected Wick contractions of the
two-point function.}
\label{fig:tadpole-graphs}
\end{figure}

The zero-momentum bridge gives this topology a simple
interpretation. The self-contracted vertex generates a homogeneous
one-loop displacement of the clock, which is then inserted into the
external two-point function.
The connected non-1PI kernels have the common form
\begin{equation}\label{eq:tadkernel-compact}
\mathcal I_{XY}^{\rm tad/1PR}
=2\mathrm{Re}\sum_{a,b=0}^{1}
\mathcal R_{X,a}\mathcal P_{Y,b}\Delta\mathcal W_{ab},
\qquad X,Y\in\{T,S\}.
\end{equation}
Here $\mathcal R_{X,a}$, $\mathcal P_{Y,b}$, and
$\Delta\mathcal W_{ab}$ denote the external-leg factors, the
self-contracted loop, and the zero-momentum bridge, respectively;
the indices $a,b$ count derivatives on the two ends of the bridge:
\begin{equation}\label{eq:tadbridge}
\mathcal W_{ab}=\partial_{t_1}^a\partial_{t_2}^bW_0(t_1,t_2),
\qquad
\Delta\mathcal W_{ab}=\mathcal W_{ab}^*-\mathcal W_{ab}.
\end{equation}
With $E_0,E_1$ evaluated at $t_1$ and $L_{ab}$ at $(k,t_2)$,
the vertex factors are
\begin{align}\label{eq:tadfactors}
\mathcal R_{T,0}&=2g_{T1}E_1^2,&
\mathcal R_{T,1}&=4g_{T1}E_0E_1,&
\mathcal R_{S,0}&=2g_{S1}p^2E_0^2,&
\mathcal R_{S,1}&=0,\\
\mathcal P_{T,0}&=g_{T2}L_{11},&
\mathcal P_{T,1}&=2g_{T2}L_{01},&
\mathcal P_{S,0}&=g_{S2}k^2L_{00},&
\mathcal P_{S,1}&=0.
\end{align}
Appendix~\ref{app:wick} gives the Wick-contraction counting.

\Needspace{12\baselineskip}
\section{Diagrammatic cancellations and momentum boundaries}\label{sec:cancellation}
We now take the large-scale limit of the finite-momentum
expressions in section~\ref{sec:bare} and derive the central
result: the sum of the derivative-vertex contributions reduces
to a momentum boundary. After specifying the external response
and the mode identity, we track the fish--quartic cancellation
and then combine the remaining local terms with the tadpole and
counterterm contributions. Keeping the momentum total derivative
explicit separates the diagrammatic cancellation from the
prescription needed to obtain a vanishing correction.

\subsection{Long-wavelength response and the dilation identity}
\label{sec:assumptions}\label{sec:modeidentity}
We take $p\to0$ with $k\gg p$ and isolate the term proportional
to the external tree soft pole $\Ptree$. The external mode is the
adiabatic long mode that left the horizon before the transient
non-slow-roll evolution. This choice selects the physical mode
associated with the dilation~\cite{HuiEtAl2019}. Accordingly,
$F_p(t,t_j)=\Ptree+\order(p^2\Ptree)$ and
$\partial_{t_j}F_p=\order(p^2\Ptree)$, whereas
\begin{equation}\label{eq:response}
R_0(t,t_j)=-\frac12\int_{t_j}^{t}\frac{\dd s}{C(s)},\qquad
C_j\partial_{t_j}R_0(t,t_j)=\frac12.
\end{equation}
Derivatives on an external \emph{retarded} line must therefore be
retained. This soft limit does not impose an adiabatic or
superhorizon approximation on the internal modes.

The single-field squeezed-limit consistency relation provides the
historical setting for the dilation response~\cite{CreminelliZaldarriaga2004}.
Previous loop analyses have used dilation consistency relations to
organize the large-scale response and its cancellations
\cite{Fumagalli2025,EmaEtAl2026}; a related mode-equation identity
in flat gauge appears in Ref.~\cite{FangGuo2026}. We therefore give
the corresponding identity explicitly in the Eulerian Goldstone EFT.
Starting from the general mode equation~\eqref{eq:mode}, we
differentiate it in time and momentum. Combining the two responses
cancels their spatial-gradient sources. The resulting forced equation for
$X_k=\dot u_k+H(\partial_{\ln k}+3/2)u_k$ has a retarded solution
fixed by the initial state in section~\ref{sec:free}. The
factor $3/2$ is the scaling weight of a normalized Fourier mode.
For the flat-slicing mode $v_k=\Pi u_k$, the clock map gives
$[\partial_t+H(\partial_{\ln k}+3/2)]v_k
=\Pi(X_k+\alpha u_k/2)$, connecting this response to the
flat-variable realization of the dilation in
section~\ref{sec:slicing-dilation}.
Appendix~\ref{app:dilation} gives the derivation and its extension
to general initial data.

To state the identity in the form needed for the loop calculation,
define, for $t_2<t_1$,
\begin{equation}\label{eq:history-kernel}
Z_k=F_k(t_1,t_2)R_k(t_1,t_2)=\mathrm{Im}(W_k^2),\qquad
h_k=C_2\partial_{t_2}Z_k,
\end{equation}
and the history integral
\begin{equation}\label{eq:Ymain}
Y_k(t_1)=\int_{t_i}^{t_1}\dd t_2\,\dot\alpha_2h_k(t_1,t_2)
=2\,\mathrm{Im}\!\left[
u_k^2(t_1)\int_{t_i}^{t_1}\dd t_2\,
C_2\dot\alpha_2u_k^*(t_2)\dot u_k^*(t_2)\right].
\end{equation}
Thus the history integral is fixed by equal-time quantities:
\begin{equation}\label{eq:Ymain-identity}
Y_k=L_{01}+\frac H2(\partial_{\ln k}+3)L_{00}
+\frac\alpha2L_{00}.
\end{equation}
This equation replaces the nonlocal fish history by
equal-time contractions without specifying a mode-function solution.
The result can then be compared directly with local contributions.

\subsection{Fish--quartic cancellation and the history integral}
\label{sec:soft}\label{sec:history}
The free mode equation simplifies the fish graphs before
the dilation identity is used. Integration over the earlier vertex
will separate an equal-time endpoint from a history term. The quartic
interaction matches the endpoint. The history term is reduced with
Eq.~\eqref{eq:Ymain-identity}.

We keep the leading external
soft pole and take $t_2<t_1$, with $t_2$ the earlier vertex.
\referencechange{Within this soft projection, the free mode equation relates
the two cubic couplings acting on hard propagators. For two kernels
$X,Y$ satisfying Eq.~\eqref{eq:mode} at a common vertex time $t_j$,
\begin{equation}\label{eq:vertexlemma}
2g_{Tj}(\partial_{t_j}X)(\partial_{t_j}Y)
+2g_{Sj}k^2XY
=-\alpha_j\partial_{t_j}
\left[C_j\partial_{t_j}(XY)\right].
\end{equation}
The two derivative couplings therefore act as a single differential
operator on the hard propagator product.
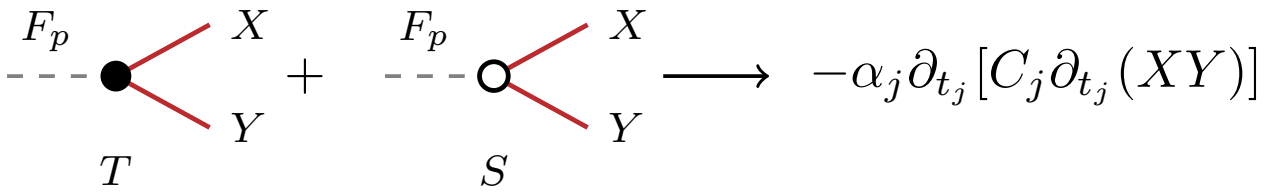
\begin{center}
\begin{tikzpicture}[baseline=0pt,font=\small,text=black]
  \draw[gray,dashed,line width=0.7pt] (-0.72,0)--(0,0);
  \node[above,font=\scriptsize] at (-0.46,0.03) {$F_p$};
  \draw[fprop] (0,0)--(0.62,0.34);
  \draw[fprop] (0,0)--(0.62,-0.34);
  \node[right,font=\scriptsize] at (0.62,0.34) {$X$};
  \node[right,font=\scriptsize] at (0.62,-0.34) {$Y$};
  \node[tvertex] at (0,0) {};
  \node[below,font=\scriptsize] at (0,-0.39) {$T$};
  \node at (1.25,0) {$+$};
  \draw[gray,dashed,line width=0.7pt] (1.78,0)--(2.5,0);
  \node[above,font=\scriptsize] at (2.04,0.03) {$F_p$};
  \draw[fprop] (2.5,0)--(3.12,0.34);
  \draw[fprop] (2.5,0)--(3.12,-0.34);
  \node[right,font=\scriptsize] at (3.12,0.34) {$X$};
  \node[right,font=\scriptsize] at (3.12,-0.34) {$Y$};
  \node[svertex] at (2.5,0) {};
  \node[below,font=\scriptsize] at (2.5,-0.39) {$S$};
  \draw[->,black,line width=0.7pt] (3.62,0)--(4.32,0);
  \node[anchor=west] at (4.45,0)
    {$-\alpha_j\partial_{t_j}[C_j\partial_{t_j}(XY)]$};
\end{tikzpicture}\\[-1pt]
{\footnotesize Earlier vertex: leading soft statistical projection.}
\end{center}}
\referencechange{Using $F_k(t_1,t_2)$ and $R_k(t_1,t_2)$, the earlier vertex gives
\begin{equation}\label{eq:Qdef}
Q_k=-\alpha_2\partial_{t_2}
\left[C_2\partial_{t_2}
\bigl(F_k(t_1,t_2)R_k(t_1,t_2)\bigr)\right]
=-\alpha_2\partial_{t_2}h_k.
\end{equation}}
The later vertex acts on $Q_k$ and the external retarded line.
Keeping derivatives on both, the sum of the fish graphs becomes
\begin{equation}\label{eq:composite}
\delta P_{\rm fish}^{\rm IR}
=2\Ptree\int_{t_i}^{t}\dd t_1
\int_{t_i}^{t_1}\dd t_2\intk
-\alpha_1\partial_{t_1}
\bigl[C_1\partial_{t_1}(R_0(t,t_1)Q_k(t_1,t_2))\bigr].
\end{equation}
\referencechange{The four ordered fish contributions are thus collected into
a single differential structure. Derivatives acting on the external
retarded line at the later vertex must still be retained.}
We now integrate by parts in $t_2$. Since $Q_k$ contains a
$t_2$ derivative, this separates an upper endpoint from the
remaining time integral:
\begin{equation}\label{eq:fishparts}
\begin{aligned}
\delta P_{\rm fish}^{\rm IR}
=2\Ptree\int_{t_i}^{t}\dd t_1\intk\alpha_1\bigg\{&
\left[\alpha_2\partial_{t_1}
\bigl(C_1\partial_{t_1}(R_0h_k)\bigr)\right]_{t_2=t_1^-}\\
&-\int_{t_i}^{t_1}\dd t_2\,\dot\alpha_2\partial_{t_1}
\bigl(C_1\partial_{t_1}(R_0h_k)\bigr)\bigg\}.
\end{aligned}
\end{equation}
The first term is evaluated at $t_2=t_1^-$; the lower endpoint is
absent with the initial-state prescription of section~\ref{sec:free}.
The second term is the remaining bulk integral, denoted by
$\delta P_{\rm fish,bulk}^{\rm IR}$.

The free equation and the Wronskian give the required equal-time values:
\begin{equation}\label{eq:main-equal-time}
\left.h_k\right|_{t_2=t_1^-}=\frac12L_{00},\qquad
\left.\partial_{t_1}h_k\right|_{t_2=t_1^-}=0,\qquad
\left.\partial_{t_1}(C_1\partial_{t_1}h_k)\right|_{t_2=t_1^-}
=-C_1(L_{11}+\omega_k^2L_{00}).
\end{equation}
Using these equal-time values in the first term of
Eq.~\eqref{eq:fishparts} gives a single $t_1$ integral:
\begin{equation}\label{eq:fishupper}
\delta P_{\rm fish,upper}^{\rm IR}
=-2\Ptree\int_{t_i}^{t}\dd t_1\,R_0(t,t_1)
\intk C_1\alpha_1^2
\left(L_{11}+\omega_k^2L_{00}\right).
\end{equation}
The quartic kernels in Eq.~\eqref{eq:quartickernels}, taken in the
same soft limit, give
\begin{align}\label{eq:quarticsoft}
\delta P_{4K}^{\rm IR}
&=4\Ptree\int_{t_i}^{t}\dd t_1\intk
R_0\lambda_KL_{11}
+8\Ptree\int_{t_i}^{t}\dd t_1\intk
(\partial_{t_1}R_0)\lambda_KL_{01},\\
\delta P_{4S}^{\rm IR}
&=4\Ptree\int_{t_i}^{t}\dd t_1\intk
R_0\lambda_Sk^2L_{00},\label{eq:quarticsoftS}
\end{align}
The mixed contraction in the first line survives because
$C_1\partial_{t_1}R_0=1/2$.

For brevity, define
\begin{equation}\label{eq:SDdefs}
\mathcal S_k=L_{11}+\omega_k^2L_{00},\qquad
\mathcal D_k=L_{11}-\omega_k^2L_{00}.
\end{equation}
Inserting the two quartic coefficients from Eq.~\eqref{eq:coeffs} into
Eqs.~\eqref{eq:quarticsoft} and \eqref{eq:quarticsoftS} gives the
complete quartic result
\begin{equation}\label{eq:quarticsummed}
\delta P_4^{\rm IR}
=2\Ptree\int_{t_i}^{t}\dd t_1\intk\left[
C_1R_0\alpha_1^2\mathcal S_k
-C_1R_0\dot\alpha_1\mathcal D_k
+(\alpha_1^2-\dot\alpha_1)L_{01}\right].
\end{equation}
The first term in Eq.~\eqref{eq:quarticsummed} cancels the fish
endpoint in Eq.~\eqref{eq:fishupper}:
\begin{equation}\label{eq:firstcancel}
\delta P_{\rm fish,upper}^{\rm IR}
+2\Ptree\int_{t_i}^{t}\dd t_1\intk
C_1R_0\alpha_1^2\mathcal S_k=0.
\end{equation}
The equal-time endpoint of the two-cubic-vertex graph is
therefore completed by the quartic Hamiltonian contribution. The
Legendre contact in $\lambda_K$ is essential to this matching.

We next turn to the bulk term in Eq.~\eqref{eq:fishparts}.
Expanding the later vertex and using
$\partial_{t_1}(C_1\partial_{t_1}R_0)=0$ gives
\begin{equation}\label{eq:fishbulk-beforeY}
\begin{aligned}
\frac{\delta P_{\rm fish,bulk}^{\rm IR}}{2\Ptree}
&=-\int_{t_i}^{t}\dd t_1\intk\alpha_1
\int_{t_i}^{t_1}\dd t_2\,\dot\alpha_2
\partial_{t_1}\!\left[C_1\partial_{t_1}(R_0h_k)\right]\\
&=-\int_{t_i}^{t}\dd t_1\intk\alpha_1
\int_{t_i}^{t_1}\dd t_2\,\dot\alpha_2
\left[R_0\partial_{t_1}(C_1\partial_{t_1}h_k)
+2C_1(\partial_{t_1}R_0)(\partial_{t_1}h_k)\right].
\end{aligned}
\end{equation}
Both terms contain the $t_2$ integral that defines $Y_k$ in
Eq.~\eqref{eq:Ymain}. Its upper limit depends on $t_1$, so
differentiating $Y_k$ also produces an equal-time term. By
Eq.~\eqref{eq:main-equal-time},
\begin{equation}\label{eq:Ymain-leibniz}
\begin{aligned}
\int_{t_i}^{t_1}\dd t_2\,\dot\alpha_2\partial_{t_1}h_k
&=\dot Y_k-\tfrac12\dot\alpha_1L_{00},\\
\int_{t_i}^{t_1}\dd t_2\,\dot\alpha_2
\partial_{t_1}(C_1\partial_{t_1}h_k)
&=\partial_{t_1}\!\left[C_1
\left(\dot Y_k-\tfrac12\dot\alpha_1L_{00}\right)\right].
\end{aligned}
\end{equation}
The bulk contribution can therefore be written as
\begin{equation}\label{eq:fishbulk-Y}
\begin{aligned}
\frac{\delta P_{\rm fish,bulk}^{\rm IR}}{2\Ptree}
=-\int_{t_i}^{t}\dd t_1\intk\alpha_1\bigg\{&
R_0\partial_{t_1}\!\left[C_1
\left(\dot Y_k-\tfrac12\dot\alpha_1L_{00}\right)\right]\\
&+2C_1(\partial_{t_1}R_0)
\left(\dot Y_k-\tfrac12\dot\alpha_1L_{00}\right)\bigg\}.
\end{aligned}
\end{equation}
We can now use Eq.~\eqref{eq:Ymain-identity}. The free mode equation
also gives
\begin{equation}\label{eq:L-eom-main}
\dot L_{00}=2L_{01},\qquad
\dot L_{01}=\mathcal D_k-(3H+\alpha)L_{01}.
\end{equation}
Define
\begin{equation}\label{eq:Bdefs}
\mathcal B_k=\mathcal D_k+H\partial_{\ln k}L_{01}.
\end{equation}
Together with Eq.~\eqref{eq:Ymain-identity}, these relations give
\begin{equation}\label{eq:Y-to-B}
\begin{aligned}
\dot Y_k-\tfrac12\dot\alpha L_{00}
&=\dot L_{01}+H(\partial_{\ln k}+3)L_{01}+\alpha L_{01}\\
&=\mathcal D_k+H\partial_{\ln k}L_{01}=\mathcal B_k.
\end{aligned}
\end{equation}
The equal-time term from the moving upper limit is canceled by
the corresponding term in the derivative of
Eq.~\eqref{eq:Ymain-identity}. Substituting Eq.~\eqref{eq:Y-to-B}
then reduces the fish bulk to a single time integral:
\begin{equation}\label{eq:fishbulkbody}
\delta P_{\rm fish,bulk}^{\rm IR}
=2\Ptree\int_{t_i}^{t}\dd t_1\intk\left[
-\alpha_1R_0\partial_{t_1}(C_1\mathcal B_k)
-2C_1\alpha_1(\partial_{t_1}R_0)\mathcal B_k\right].
\end{equation}
After the endpoint cancellation, the remaining quartic contribution is
\begin{equation}\label{eq:quarticrembody}
\delta P_{4,\rm rem}^{\rm IR}
=2\Ptree\int_{t_i}^{t}\dd t_1\intk\left[
-C_1R_0\dot\alpha_1\mathcal D_k
+(\alpha_1^2-\dot\alpha_1)L_{01}\right].
\end{equation}
Adding Eqs.~\eqref{eq:fishbulkbody} and
\eqref{eq:quarticrembody} gives
\begin{equation}\label{eq:fishquartic}
\begin{aligned}
\frac{\delta P_{\rm fish+4}^{\rm IR}}{2\Ptree}
&=\frac{\delta P_{\rm fish,bulk}^{\rm IR}
+\delta P_{4,\rm rem}^{\rm IR}}{2\Ptree}\\
&=\int_{t_i}^{t}\dd t_1\intk\left[
-\frac{\alpha_1}{2}\mathcal B_k
+C_1R_0(t,t_1)\dot\alpha_1H_1\partial_{\ln k}L_{01}
+(\alpha_1^2-\dot\alpha_1)L_{01}\right].
\end{aligned}
\end{equation}
To obtain the second line, integrate the first term of
Eq.~\eqref{eq:fishbulkbody} by parts in $t_1$. Its boundary terms
vanish because $R_0(t,t)=0$ and $\alpha_i=0$. The terms involving
$\mathcal B_k$ without $R_0$ combine to $-\alpha_1\mathcal B_k/2$,
while the terms proportional
to $C_1R_0\dot\alpha_1$ give
$C_1R_0\dot\alpha_1(\mathcal B_k-\mathcal D_k)
=C_1R_0\dot\alpha_1H_1\partial_{\ln k}L_{01}$.
The remaining local terms will be combined with the tadpoles and
counterterms in the next subsection.

\Needspace{12\baselineskip}
\subsection{Tadpoles, counterterms, and the momentum boundary}
\label{sec:ct}\label{sec:boundary}
After the fish--quartic cancellation, a local response of
the external long mode remains. Part of this response is associated
with the homogeneous displacement generated by the hard loop. The
non-1PI tadpole makes this background response explicit. Its
renormalization is fixed by the one-point condition.

We now collect the equal-time loop integrals. Time derivatives below
act on the integrands at fixed loop momentum. Define
\begin{align}\label{eq:JMKSigma}
J&=\intk L_{01},& M&=\intk\partial_{\ln k}L_{01},\\
K_{\partial k}&=M+3J
=\intk(\partial_{\ln k}+3)L_{01},
&\Sigma_\pi&=\intk[\alpha\mathcal D_k+2\dot\alpha L_{01}].
\end{align}

The direct non-1PI contractions reduce to a zero-mode bridge
$\pi_{\rm 1L}^{\rm tad}$ and a bridge differentiated at its later
vertex, $(\dot\pi)_{\rm 1L}^{\rm tad}$.  Their soft-limit expressions
are
\begin{equation}\label{eq:tad-main-bridge}
\begin{aligned}
\pi_{\rm 1L}^{\rm tad}(t_1)
&=\int_{t_i}^{t_1}\dd t_2\,C_2R_0(t_1,t_2)\Sigma_\pi(t_2),\\
(\dot\pi)_{\rm 1L}^{\rm tad}(t_1)
&=\partial_{t_1}\pi_{\rm 1L}^{\rm tad}(t_1)+\alpha_1J_1.
\end{aligned}
\end{equation}
The $\alpha_1J_1$ term is the equal-time upper endpoint of the
differentiated bridge.  Consequently the two-point tadpole is
\begin{equation}\label{eq:tad-main-split}
\frac{\delta P_{\rm tad/1PR}^{\rm IR}}{2\Ptree}
=-\int_{t_i}^{t}\dd t_1\,\alpha_1
\partial_{t_1}\pi_{\rm 1L}^{\rm tad}(t_1)
-\int_{t_i}^{t}\dd t_1\,\alpha_1^2J_1.
\end{equation}
The direct Wick contractions, endpoint identity, and counterterm
relation used below are derived in App.~\ref{app:tad}.

The one-point condition $\langle\pi\rangle=0$ fixes the linear
counterterm $a^3b_1\pi$ by matching the same retarded source:
\begin{equation}\label{eq:tad-main-ct}
\begin{aligned}
a^3b_1&=C\Sigma_\pi,\\
\pi_{\rm ct}(t_1)
&=-\int_{t_i}^{t_1}\dd t_2\,C_2R_0(t_1,t_2)\Sigma_\pi(t_2)
=-\pi_{\rm 1L}^{\rm tad}(t_1).
\end{aligned}
\end{equation}
\referencechange{The one-point condition removes the propagating tadpole response.
The contact from the differentiated bridge survives:
\begin{equation}\label{eq:tad-contact-after-ct}
\begin{aligned}
\pi_{\rm 1L}^{\rm tad}+\pi_{\rm ct}&=0,\\
(\dot\pi)_{\rm 1L}^{\rm tad}
+(\dot\pi)_{\rm ct}&=\alpha J.
\end{aligned}
\end{equation}
Here $(\dot\pi)_{\rm ct}=\partial_t\pi_{\rm ct}$.
The second line is not the time derivative of the first: the bridge
derivative acts at fixed integration domain. Its contribution to the
two-point function is}
\begin{equation}\label{eq:tad-main-rem}
\frac{\delta P_{\rm tad/1PR}^{\rm IR}
+\delta P_{b_1}^{\rm IR}}{2\Ptree}
=-\int_{t_i}^{t}\dd t_1\,\alpha_1^2J_1.
\end{equation}
\referencechange{The surviving contact is essential to the two-point cancellation.
It removes the $+\alpha^2J$ term from the mixed quartic contraction
in Eq.~\eqref{eq:fishquartic}.} The $-\dot\alpha J$
term and the fish-history terms remain for the subsequent
reduction. We obtain the linear and quadratic counterterms by
expanding a single counterterm potential, which fixes
$b_2=\dot b_1/2$. The quadratic coefficient is therefore determined
by the one-point condition rather than adjusted independently
\cite{FangGuo2026,Inomata2025Counterterms}.
Its two-point insertion is
\begin{equation}\label{eq:tad-main-b2}
\frac{\delta P_{b_2}^{\rm IR}}{2\Ptree}
=-\int_{t_i}^{t}\dd t_1\,C_1R_0(t,t_1)
(\dot\Sigma_{\pi,1}+\alpha_1\Sigma_{\pi,1}).
\end{equation}
Combining these terms with Eq.~\eqref{eq:fishquartic} and setting
$B=\intk\mathcal B_k$ gives
\begin{equation}\label{eq:combined-loop-ct}
\frac{\delta P_{\rm loop+ct}^{\rm IR}}{2\Ptree}
=\int_{t_i}^{t}\dd t_1\left\{
-\frac{\alpha_1}{2}B_1-\dot\alpha_1J_1
+C_1R_0(t,t_1)
\bigl[\dot\alpha_1H_1M_1-\dot\Sigma_{\pi,1}
-\alpha_1\Sigma_{\pi,1}\bigr]\right\}.
\end{equation}

Using the free mode equation and the zero-momentum response,
Eq.~\eqref{eq:combined-loop-ct} reduces to
\begin{equation}\label{eq:result}
\begin{aligned}
\delta P_{\rm loop+ct}^{\rm IR}(p,t)
&=2\Ptree\int_{t_i}^{t}\dd t_1\,
R_0(t,t_1)\,\mathcal W(t_1),\\
\mathcal W
&=\partial_t(C\alpha HK_{\partial k})
+CH\dot\alpha K_{\partial k}+\order(\dot H).
\end{aligned}
\end{equation}
The remaining momentum structure is a total derivative:
\begin{equation}\label{eq:momentum-total-derivative}
K_{\partial k}
=\frac{1}{2\pi^2}\int\dd k\,
\partial_k\!\left[k^3\mathrm{Re}(u_k\dot u_k^*)\right].
\end{equation}
The steps leading from Eq.~\eqref{eq:combined-loop-ct} to
Eq.~\eqref{eq:result} are given in App.~\ref{app:source-reduction}.
In particular, the terms proportional to $J$ and $\dot J$ cancel
at the order retained here.

The momentum boundary can remain ultraviolet divergent after this
counterterm cancellation: the one-point condition fixes the
tadpole subtraction, while evaluating the boundary requires a
regularization prescription. For the adiabatic ultraviolet modes,
$L_{01}=-(2H+\alpha)/(8a^2Ak)+\order(k^{-3})$, so
$k^3L_{01}$ grows quadratically at large $k$ and a fixed comoving
cutoff leaves a quadratic ultraviolet divergence in
Eq.~\eqref{eq:momentum-total-derivative}. A related divergent
momentum boundary appears in the flat-gauge calculation of
Ref.~\cite{FangGuo2026}, where Ward identities impose further
constraints on the regularized loop integrals and their
counterterms. In that analysis, the cancellation of the remaining
contribution relies on these constraints on the regularization
and subtraction procedure. In particular, a time-dependent
momentum cutoff contributes through its moving integration
surface. We therefore keep $K_{\partial k}$ explicit, leaving
its evaluation to a consistent choice of regulator and
renormalization prescription
\referencechange{\cite{SheikhahmadiNassiriRad2024,BallesterosEtAl2025Finite,BallesterosEtAl2026Finite}}.

\section{Discussion}\label{sec:discussion}
Our explicit calculation gives a rather specific statement. In the
decoupling limit considered here, the one-loop contribution after
including the counterterms fixed by the one-point condition can be
reduced to the momentum-boundary form in Eq.~\eqref{eq:result}.
This reduction follows from the free mode equation and does not
require an explicit solution for the internal modes. The
non-boundary terms cancel once the time-history contribution is
rewritten in terms of equal-time quantities. What remains is the
total derivative in Eq.~\eqref{eq:momentum-total-derivative}. We
leave its momentum integral unevaluated. The calculation therefore
establishes the cancellation of the non-boundary part, rather than
a regulator-independent vanishing theorem for the full loop
correction.

The resulting structure closely parallels what has been found in
flat and comoving descriptions using symmetry and consistency
relations~\referencechange{\cite{Fumagalli2025,ArtigasEtAl2026,FangGuo2026,EmaEtAl2026}}. The
nonlinear maps derived here provide a simple way to understand why
these calculations can look very different while probing the same
adiabatic long mode. In flat slicing, the fixed-background
nonlinearities are carried by the scalar potential. The clock
redefinition expresses the same background time dependence through
derivative Goldstone interactions. Individual diagrams therefore
depend strongly on the chosen field variable, while the cancellation
is organized by the same dilation symmetry. From this perspective,
the Goldstone calculation gives a diagrammatic realization of the
relation between the simpler flat-slicing interactions and the
symmetry-based results obtained in other variables.

The choice of one-point renormalization condition provides a simple
example of the dependence on field variables. Here we impose
$\langle\pi\rangle_{\rm 1L}=0$, whereas Ref.~\cite{FangGuo2026}
imposes a vanishing one-point function for the flat-slicing field.
The two conditions are not related linearly. Before renormalization
of the corresponding composite operator, the nonlinear clock map
gives
\begin{equation}\label{eq:onepoint-map}
\langle\delta\phi\rangle_{\rm 1L}
=\Pi\langle\pi\rangle_{\rm 1L}
+\frac{\dot\Pi}{2}\langle\pi^2\rangle_0,
\end{equation}
where the coincident contraction is evaluated with the same
regulator as the one-point function. Thus a condition that leaves
the Goldstone background unshifted does not, by itself, leave the
flat-field background unshifted. Comparing the two prescriptions
requires a common definition of the background split and of the
renormalized composite operators. A systematic comparison of these
choices would help clarify how the finite renormalization depends
on the field parametrization.

\referencechange{The same comparison can be extended to finite external momentum. EFT
calculations identify ultraviolet divergences with nontrivial
dependence on the external wavenumber and organize their subtraction
through local derivative
counterterms~\cite{BragliaPinol2025,BragliaPinol2026Freezing,BragliaCespedesPinol2026}.
The nonlinear clock map provides a way to compare how these terms
appear in the Goldstone and spatially flat descriptions. In the
latter, a recent analysis relaxes the hierarchy between loop and
external wavenumbers and reports a cancellation of superhorizon
time dependence after including the nonlinear relation to curvature
perturbations and the relevant boundary
terms~\cite{Inomata2026AllScales}. We will clarify the structure and
renormalization of the contributions at finite external momentum
in a subsequent paper.}

\referencechange{A further extension is to include gravitational mixing beyond
the leading scalar-gravity decoupling limit used here. This requires}
solving the
lapse and shift constraints and substituting their nonlinear
solutions back into the action. This generates spatially nonlocal
interactions and substantially complicates the loop calculation.
The decoupling limit is therefore useful not only as an
approximation, but also as a controlled setting in which the
cancellation associated with the scalar clock can be isolated
explicitly. Within the power counting adopted in
Sec.~\ref{sec:eft}, the omitted gravitational terms are suppressed
by an additional power of $\epsilon$, while the time dependence
relevant for transient USR is retained. Extending the calculation
to include the constraint sector would provide a valuable check of
how the same symmetry structure is realized once gravitational
mixing is restored.

\acknowledgments
We thank Jason Kristiano, Keisuke Inomata, Diego Cruces and Junichi Yokoyama
for useful comments.
This work is supported in part by the National Natural Science Foundation
of China (No. 12475067 and No. 12235019).

\appendix
{\color{black}
\section{Inverse clock map and comoving slicing}
\label{app:inverse-clock-map}
The inverse map of Eq.~\eqref{eq:inverse-clock} is
\begin{equation}
t_1=t+\pi(t,\mathbf x),\qquad
t=t_1-\pi_1(t_1,\mathbf x),\qquad
\pi_1(t_1,\mathbf x)=\pi(t,\mathbf x).
\end{equation}
A dot on $\pi_1$ denotes differentiation with respect to $t_1$ at
fixed $\mathbf x$. Differentiating the inverse relation at fixed
$\mathbf x$ gives $\partial t/\partial t_1=1-\dot{\pi}_1$.
The direct clock relation gives $\partial t_1/\partial t=1+\dot\pi$.
These derivatives are reciprocal, so
\begin{equation}\label{eq:inverse-clock-derivatives}
\frac{\partial t}{\partial t_1}=1-\dot{\pi}_1,
\qquad 1+\dot\pi=\frac{1}{1-\dot{\pi}_1}.
\end{equation}
At fixed old time $t$, differentiating
$t=t_1-\pi_1(t_1,\mathbf x)$ with respect to $x^i$ yields
\begin{equation}
0=(1-\dot{\pi}_1)(\partial_i t_1)_t
-(\partial_i\pi_1)_{t_1},
\qquad
(\partial_i\pi)_t=(\partial_i t_1)_t
=\frac{(\partial_i\pi_1)_{t_1}}{1-\dot{\pi}_1}.
\end{equation}
The subscripts distinguish the time held fixed in each spatial
derivative.

Starting from the exact Eulerian action
\eqref{eq:eulerian-exact}, substitute
$t=t_1-\pi_1$ and $t+\pi=t_1$. The time measure becomes
$\dd t=(1-\dot{\pi}_1)\dd t_1$, while its kinetic bracket becomes
\begin{equation}
(1+\dot\pi)^2-\frac{(\partial_i\pi)^2}{a^2(t)}
=\frac{1}{(1-\dot{\pi}_1)^2}
\left[1-\frac{(\partial_i\pi_1)^2}{a^2(t_1-\pi_1)}\right].
\end{equation}
The same change of variables therefore gives
\begin{align}
S[\pi_1]=\int\dd t_1\,\dd^3x\,a^3(t_1-\pi_1)
\bigg\{&\frac{\Pi^2(t_1)}{2(1-\dot{\pi}_1)}
\left[1-\frac{(\partial_i\pi_1)^2}{a^2(t_1-\pi_1)}\right]
\nonumber\\
&-(1-\dot{\pi}_1)V\!\left(\bar\phi(t_1)\right)\bigg\},
\end{align}
as quoted in Eq.~\eqref{eq:inverse-action}. No background equation
or integration by parts is used in this substitution.

For the large dilation, write $\Delta(t)=\lambda/H(t)$ and take
\begin{equation}
\widetilde{\mathbf x}=(1-\lambda)\mathbf x,\qquad
\widetilde t=t+\Delta(t),\qquad
\widetilde\pi(\widetilde t,\widetilde{\mathbf x})
=\pi(t,\mathbf x)-\Delta(t).
\end{equation}
The clock is invariant:
$\widetilde t+\widetilde\pi=t+\pi=t_1$. In the inverse map this
also gives
\begin{equation}
\widetilde t=t_1-\tilde{\pi}_1
=t_1-\pi_1+\Delta(t),
\qquad
\tilde{\pi}_1(t_1,\widetilde{\mathbf x})
=\pi_1(t_1,\mathbf x)-\Delta(t)
\simeq\pi_1(t_1,\mathbf x)-\frac{\lambda}{H(t_1)}.
\end{equation}
The last equality uses the order retained in
Sec.~\ref{sec:eft}.

Write the transformed scale factor as
$\widetilde a(t+\Delta)=a(t)/(1-\lambda)$ at this order.
Since $\dd^3\widetilde x\simeq(1-3\lambda)\dd^3x$, the shifted
measure satisfies
\begin{equation}\label{eq:inverse-dilation-factors}
a^3(t_1-\pi_1)\,\dd^3x
\simeq\widetilde a^3(t_1-\tilde{\pi}_1)\,
\dd^3\widetilde x.
\end{equation}
The time derivative of $\Delta$ is beyond the retained order.
Consequently the velocity factor is unchanged, and the spatial
gradient has the same normalization:
\begin{equation}
1-\dot{\pi}_1\simeq1-\dot{\tilde{\pi}}_1,
\qquad
\frac{\partial_i\pi_1}{a(t_1-\pi_1)}
\simeq\frac{\widetilde\partial_i\tilde{\pi}_1}
{\widetilde a(t_1-\tilde{\pi}_1)}.
\end{equation}
The coefficients $\Pi(t_1)$ and $V(\bar\phi(t_1))$ are unchanged
because $t_1$ is fixed. These relations give
$S[\tilde{\pi}_1]\simeq S[\pi_1]$ without repeating the full action.

These relations justify the compact fixed-$t_1$ symmetry statement
used in Sec.~\ref{sec:slicing-dilation}. The identification with
$\zeta$ follows from the induced spatial geometry discussed there.
\par}

\section{Interaction Hamiltonian}\label{app:ham}
The momentum follows by differentiating the velocity term of the
\emph{full} bulk Lagrangian in Eq.~\eqref{eq:action}:
\begin{equation}\label{eq:AppHam}
P_\pi\equiv\frac{\partial\mathcal L_{\rm bulk}}{\partial\dot\pi}
=2a^3K\dot\pi
=2a^3A\dot\pi+2a^3\dot A\pi\dot\pi
+a^3\ddot A\pi^2\dot\pi.
\end{equation}
The two interaction terms in $P_\pi$ contribute to the cubic and
quartic Hamiltonians. Solving Eq.~\eqref{eq:AppHam} for the velocity
and performing the
Legendre transform while $P_\pi$ is an independent phase-space
variable gives
\begin{equation}\label{eq:fullham}
\dot\pi=\frac{P_\pi}{2a^3K},\qquad
\mathcal H_{\rm bulk}=P_\pi\dot\pi-\mathcal L_{\rm bulk}
=\frac{P_\pi^2}{4a^3K}+aK(\partial_i\pi)^2.
\end{equation}
The kinetic term contains $1/K$, so its expansion is the next
necessary step.  Write
$K=A(1+x)$ with
$x=(\dot A/A)\pi+(\ddot A/2A)\pi^2$.
Keeping $1-x+x^2$ through quadratic order in $\pi$ yields
\begin{equation}\label{eq:inverseK}
\frac1K=\frac1A\left[1-\frac{\dot A}{A}\pi
+\left(\frac{\dot A^2}{A^2}-\frac{\ddot A}{2A}\right)\pi^2
\right]+\order(\pi^3)
\end{equation}
and hence the canonical Hamiltonian, grouped by field order, is
\begin{align}\label{eq:H234phase}
\mathcal H_2&=\frac{P_\pi^2}{4a^3A}+aA(\partial_i\pi)^2,\\
\mathcal H_3&=-\frac{\dot A}{A}\pi\frac{P_\pi^2}{4a^3A}
+a\dot A\pi(\partial_i\pi)^2,\\
\mathcal H_4&=\left(\frac{\dot A^2}{A^2}-\frac{\ddot A}{2A}\right)
\pi^2\frac{P_\pi^2}{4a^3A}
+\frac{a\ddot A}{2}\pi^2(\partial_i\pi)^2.
\end{align}
The term $\dot A^2\pi^2/A^2$ in $\mathcal H_4$ is the square of the
cubic part of $x$, generated by the Legendre transform.

Equation~\eqref{eq:H234phase} is a phase-space expansion.  Now define
the interaction picture using $H_2$.  Its Hamilton equation gives
$\dot\pi_I=P_{\pi I}/(2a^3A)$, so that
$P_{\pi I}=2a^3A\dot\pi_I$. We then express the $H_3$ and $H_4$
phase-space terms in free-field time derivatives. The
velocity-dependent terms become
\begin{equation}
\begin{aligned}
\frac{P_{\pi I}^2}{4a^3K(\pi_I,t)}
={}&a^3A\dot\pi_I^2-a^3\dot A\pi_I\dot\pi_I^2\\
&
+a^3\left(\frac{\dot A^2}{A}-\frac{\ddot A}{2}\right)
\pi_I^2\dot\pi_I^2+\cdots.
\end{aligned}
\end{equation}
Reading off the coefficients gives Eq.~\eqref{eq:coeffs}.
The Legendre transform contributes $C\alpha^2$ to the kinetic
quartic coefficient.
In a configuration-space path integral this Legendre contact is represented by the
equal-time term arising when derivatives hit the step functions in
time-ordered propagators.  Mixed coincident products are taken in
the Weyl/midpoint prescription; changing ordering would require a
corresponding change of local contact terms.

\section{Wick combinatorics and derivative placement}\label{app:wick}
The finite-momentum kernels and their integration measures are
given in section~\ref{sec:bare}. Here we spell out the
multiplicities and derivative placements entering those results.

For the equal-time insertion
$Q=\pi_{\mathbf p}(t)\pi_{-\mathbf p}(t)$, the one-vertex term is
\begin{equation}
i\int\dd t_1\langle[H_4(t_1),Q]\rangle
=2\int\dd t_1\,\mathrm{Im}\langle QH_4(t_1)\rangle.
\end{equation}
At second order, with $t_2<t_1$,
\begin{equation}\label{eq:ininwick}
\begin{aligned}
&-\langle[H_3(t_2),[H_3(t_1),Q]]\rangle\\
&\qquad=2\mathrm{Re}\left[
\langle H_3(t_2)QH_3(t_1)\rangle
-\langle QH_3(t_1)H_3(t_2)\rangle\right].
\end{aligned}
\end{equation}
These expressions fix the in-in signs before Wick contractions
are performed.

The kinetic factors in Eq.~\eqref{eq:quartickernels} follow by assigning the two labelled external
legs to the four vertex fields:
\begin{center}
\begin{tabular}{@{}lll@{}}
\toprule
External attachments & Number of pairings & Remaining loop \\
\midrule
$\dot\pi,\dot\pi$ & $2$ & $L_{00}$ \\
$\pi,\pi$ & $2$ & $L_{11}$ \\
$\pi,\dot\pi$ & $2\times2\times2=8$ & $L_{01}$ \\
\bottomrule
\end{tabular}
\end{center}
The last factor $2$ in the mixed row exchanges the two labelled
external legs.  The remaining $\pi\dot\pi$ loop is symmetrized by
the Weyl rule.  For the spatial graph, attaching both external legs
to gradients gives $2p^2$, while attaching neither gives $2k^2$.
The mixed class is proportional to $\mathbf p\cdot\mathbf k$ and
vanishes under a symmetric loop integration.  The integration measure is given in Eq.~\eqref{eq:kernel-integrals}.

The fish propagator products in Eq.~\eqref{eq:fishbase}
arise, respectively, from
$\langle QH_3(t_1)H_3(t_2)\rangle$ and
$\langle H_3(t_2)QH_3(t_1)\rangle$. Using the line derivatives
defined in Eq.~\eqref{eq:Dt}, the vertex rules are
\begin{equation}
\mathbb V_T^{(j)}=2g_{Tj}\mathbb D_j,\quad
\mathbb V_S^{(j)}=g_{Sj}\mathcal K_{pkq}.
\end{equation}
For example,
\begin{equation}
\partial_{1,e}\partial_{2,k}\mathscr K_{Q12}
=[\partial_{t_1}W_p(t,t_1)]W_p(t,t_2)
[\partial_{t_2}W_k(t_1,t_2)]W_q(t_1,t_2).
\end{equation}
On $\mathscr K_{2Q1}$, the $t_2$ derivative instead acts on
$W_k^*(t_1,t_2)$ (or $W_p^*(t,t_2)$ for the $e$ line).
The $2$ in $\mathbb V_T$ exchanges the two identical
time-differentiated fields.  The spatial rule follows from
$-2g_S\sum_{a<b}\mathbf r_a\cdot\mathbf r_b
=g_S\sum_a r_a^2$ at a momentum-conserving vertex.  Exchanging
the two labelled external attachments gives $2$, while exchanging
the two internal lines gives a graph factor $1/2$; they cancel.

The product $\mathbb D_1\mathbb D_2$ contains the nine
combinations of the three derivative pairs in Eq.~\eqref{eq:Dt};
$TS$ has the three pairs at $t_1$, while $ST$ has the three
at $t_2$.

The connected non-1PI topology has the later vertex attached to
both external legs and the earlier vertex self-contracted into a
loop, joined by the zero-momentum bridge in Eq.~\eqref{eq:tadbridge}.
The factors in Eq.~\eqref{eq:tadfactors} follow by counting the
attachments at each end. At the later $T$ vertex, an undifferentiated bridge
leaves two $\dot\pi$ external attachments ($2$ pairings); a
differentiated bridge leaves one $\pi$ and one $\dot\pi$ external
attachment ($2$ choices for the bridge times $2$ assignments of
external legs).  At the earlier $T$ vertex, the corresponding loop
is respectively $L_{11}$ or $2L_{01}$.  For either $S$ vertex a
gradient on the zero-momentum bridge vanishes, leaving only the
undifferentiated-bridge entry in Eq.~\eqref{eq:tadfactors}.
The four explicit contraction classes are
\begin{align}\label{eq:tadkernels}
\mathcal I_{TT}^{\rm tad/1PR}
&=2\mathrm{Re}\sum_{a,b=0}^1
\mathcal R_{T,a}\mathcal P_{T,b}\Delta\mathcal W_{ab},\\
\mathcal I_{TS}^{\rm tad/1PR}
&=2\mathrm{Re}\sum_{a=0}^1
\mathcal R_{T,a}\mathcal P_{S,0}\Delta\mathcal W_{a0},\\
\mathcal I_{ST}^{\rm tad/1PR}
&=2\mathrm{Re}\sum_{b=0}^1
\mathcal R_{S,0}\mathcal P_{T,b}\Delta\mathcal W_{0b},\\
\mathcal I_{SS}^{\rm tad/1PR}
&=2\mathrm{Re}\,
\mathcal R_{S,0}\mathcal P_{S,0}\Delta\mathcal W_{00}.
\end{align}
The reverse time ordering, in which the later vertex contains the
self-contraction, has complex-conjugate terms inside the real
in-in difference and vanishes. Equation~\eqref{eq:tadkernels}
therefore gives the connected non-1PI sector directly from
two-point Wick pairings.

\section{Soft-limit vertex and endpoint lemmas}\label{app:softlemmas}
The finite-momentum fish expression must be projected onto the term
carrying the long-mode statistical factor.  Set $q=k$ only after
the vertex derivatives have acted.  In the ordered domain,
$W_k=F_k+iR_k/2$ and $W_k^*=F_k-iR_k/2$.  The undifferentiated
in-in difference in Eq.~\eqref{eq:fishbase} then becomes
\begin{equation}\label{eq:softWdifference}
\mathscr K_{2Q1}-\mathscr K_{Q12}
=-2iW_p(t,t_1)\left[
F_p(t,t_2)F_kR_k
+\frac{R_p(t,t_2)}2
\left(F_k^2-\frac{R_k^2}{4}\right)\right].
\end{equation}
The first bracket carries the external soft statistical pole
$F_p\simeq\Ptree$; the second contributes at subleading soft order.
At leading order the earlier vertex differentiates the short kernels,
as $\partial_{t_2}F_p=\order(p^2\Ptree)$. The later vertex acts on
an external $R_0$, whose time derivative is unsuppressed by
Eq.~\eqref{eq:response}.

For two hard line kernels $X,Y$ satisfying the free equation at
the same vertex time $t_j$, the product rule gives
\begin{equation}\label{eq:productlemma}
\partial_{t_j}\left[C_j\partial_{t_j}(XY)\right]
=2C_j\left[(\partial_{t_j}X)(\partial_{t_j}Y)
-\omega_{k,j}^2XY\right].
\end{equation}
\referencechange{Combining this product rule with $g_T=-C\alpha$ and
$g_S=C\alpha/a^2$ gives the vertex identity in
Eq.~\eqref{eq:vertexlemma}.}
At the earlier vertex, $X=F_k(t_1,t_2)$ and
$Y=R_k(t_1,t_2)$ immediately yield $Q_k$ in
Eq.~\eqref{eq:Qdef}.  At the later vertex the first term is
$-\alpha_1R_0\partial_{t_1}(C_1\partial_{t_1}Q_k)$.
The two placements of one time derivative on the external
retarded line give
$-2C_1\alpha_1(\partial_{t_1}R_0)(\partial_{t_1}Q_k)$.
Because $\partial_{t_1}(C_1\partial_{t_1}R_0)=0$ at $p=0$,
their sum is Eq.~\eqref{eq:composite}, including the required
external-response derivatives.

Substituting $Q_k=-\alpha_2\partial_{t_2}h_k$ and integrating
over the earlier time gives, explicitly,
\begin{equation}\label{eq:t2parts}
\begin{aligned}
\delta P_{\rm fish}^{\rm IR}
=2\Ptree\int_{t_i}^{t}\dd t_1\intk\alpha_1
\bigg\{&\left[\alpha_2\partial_{t_1}
\bigl(C_1\partial_{t_1}(R_0h_k)\bigr)\right]_{t_2=t_i}^{t_2=t_1^-}\\
&-\int_{t_i}^{t_1}\dd t_2\,\dot\alpha_2
\partial_{t_1}\bigl(C_1\partial_{t_1}(R_0h_k)\bigr)
\bigg\}.
\end{aligned}
\end{equation}
The first line contains the upper and initial endpoints; the second
is the historical bulk.  The partial $t_1$ derivatives are taken
\emph{before} setting $t_2=t_1^-$.

At equal time, the Wronskian and
$Z_k=\mathrm{Im}(W_k^2)$ imply Eq.~\eqref{eq:main-equal-time}.
For example,
$C_2\partial_{t_2}\mathrm{Im}(W_k^2)|_{\rm eq}
=2C L_{00}\mathrm{Im}(u_k\dot u_k^*)=L_{00}/2$.
For the last equality in Eq.~\eqref{eq:main-equal-time}, apply
Eq.~\eqref{eq:mode} to $W_k^2$ at the later time:
\begin{equation}
\partial_{t_1}\bigl(C_1\partial_{t_1}Z_k\bigr)
=2C_1\,\mathrm{Im}\!\left[
(\partial_{t_1}W_k)^2-\omega_{k,1}^2W_k^2\right].
\end{equation}
Acting with $C_2\partial_{t_2}$ and then taking equal time gives
\begin{equation}
\eval{C_2\partial_{t_2}\mathrm{Im}[(\partial_{t_1}W_k)^2]}
=-\tfrac12L_{11},\qquad
\eval{C_2\partial_{t_2}\mathrm{Im}(W_k^2)}
=\tfrac12L_{00}.
\end{equation}
These two terms produce $-C_1L_{11}$ and
$-C_1\omega_k^2L_{00}$.  The middle equality in
Eq.~\eqref{eq:main-equal-time} follows because both terms in
$\partial_{t_1}\partial_{t_2}W_k^2$ are real at equal time.
Finally, the external-response derivative in
$\partial_{t_1}[C_1\partial_{t_1}(R_0h_k)]$ multiplies
$\eval{\partial_{t_1}h_k}=0$ by the equal-time identity,
reproducing the full upper endpoint in Eq.~\eqref{eq:fishupper}.

The quartic kinetic Wick factor in Eq.~\eqref{eq:quartickernels}
must be expanded before the soft limit.  With
$E_0=F_p+iR_p/2$ and
$E_1=\partial F_p+i\partial R_p/2$,
\begin{align}\label{eq:mixedsoft}
2\mathrm{Im}(E_1^2)&=\order(p^2\Ptree),&
2\mathrm{Im}(E_0^2)&=2\Ptree R_0,\\
2\mathrm{Im}(E_0E_1)&=\Ptree\partial_{t_1}R_0
+\order(p^2\Ptree).&&
\end{align}
Thus the mixed $8E_0E_1L_{01}$ term survives and contributes
$2\Ptree\int\dd t_1(\alpha_1^2-\dot\alpha_1)\intk L_{01}$.
The other quartic terms decompose as
\begin{equation}
2(\lambda_KL_{11}+\lambda_Sk^2L_{00})
=C\alpha^2(L_{11}+\omega_k^2L_{00})
-C\dot\alpha(L_{11}-\omega_k^2L_{00}),
\end{equation}
which proves the first cancellation in Eq.~\eqref{eq:firstcancel}.

\section{Dilation identity, initial data, and the fish history}\label{app:dilation}
\subsection{Mode-equation derivation and initial data}\label{app:dilationproof}
We derive the identity quoted in section~\ref{sec:modeidentity}
without choosing a particular mode-function solution.
Introduce the differential operators
\begin{equation}\label{eq:mode-operators}
\mathscr M_k=\partial_{\ln k},\qquad
\mathcal E_k=\partial_t^2+(3H+\alpha)\partial_t+\omega_k^2,
\qquad \mathcal E_ku_k=0.
\end{equation}
Differentiating this equation in time and in momentum, with
$\dot\omega_k^2=-2H\omega_k^2$ and
$\mathscr M_k\omega_k^2=2\omega_k^2$, gives
\begin{align}\label{eq:mode-differentiated}
\mathcal E_k\dot u_k
&=2H\omega_k^2u_k-(3\dot H+\dot\alpha)\dot u_k,\\
\mathcal E_k(\mathscr M_ku_k)&=-2\omega_k^2u_k.
\end{align}
At the decoupling order specified in Sec.~\ref{sec:eft},
the two gradient sources cancel in the combination
\begin{equation}\label{eq:Xforced}
X_k\equiv\dot u_k+H(\mathscr M_k+\tfrac32)u_k,
\qquad \mathcal E_kX_k=-\dot\alpha\dot u_k.
\end{equation}
The $3/2$ is fixed by the scaling weight of the normalized
Fourier mode. Dilation covariance of the Bunch--Davies modes in
the initial scaling phase specified in section~\ref{sec:free}
sets $X_k(t_i)=\dot X_k(t_i)=0$.
With the sign convention in Eq.~\eqref{eq:mode},
$\mathcal E_{k,t_1}R_k(t_1,t_2)=-\delta(t_1-t_2)/(2C_2)$,
so the unique retarded response is
\begin{equation}\label{eq:Xretarded}
X_k(t_1)=2\int_{t_i}^{t_1}\dd t_2\,
C_2R_k(t_1,t_2)\dot\alpha_2\dot u_k(t_2).
\end{equation}

Using the definitions of $Z_k$, $h_k$, and $Y_k$ in
Eqs.~\eqref{eq:history-kernel} and \eqref{eq:Ymain}, we now relate
this response to the two-line loop kernel.
Taking the real part of $u_kX_k^*$ in
Eq.~\eqref{eq:Xretarded} and using
$C\,\mathrm{Im}(u_k\dot u_k^*)=1/4$ yields
\begin{equation}\label{eq:YX-main}
Y_k=\mathrm{Re}(u_kX_k^*)+\frac{\alpha}{2}L_{00},\qquad
\mathrm{Re}(u_kX_k^*)=L_{01}
+\frac H2(\partial_{\ln k}+3)L_{00}.
\end{equation}
Combining these relations proves Eq.~\eqref{eq:Ymain-identity}.

The differential identity and its initial-state input can be
separated. For general finite initial data,
Eq.~\eqref{eq:Xforced} still holds at the retained order, but its
solution contains a homogeneous term:
\begin{equation}\label{eq:Xhom}
\begin{aligned}
X_k(t_1)&=X_k^{\rm hom}(t_1)
+2\int_{t_i}^{t_1}\dd t_2\,C_2R_k(t_1,t_2)
\dot\alpha_2\dot u_k(t_2),\\
\mathcal E_kX_k^{\rm hom}&=0,\qquad
X_k^{\rm hom}(t_i)=X_k(t_i),\qquad
\dot X_k^{\rm hom}(t_i)=\dot X_k(t_i).
\end{aligned}
\end{equation}
Both initial data are required. Adiabaticity of the external long
mode alone does not set them to zero. Inserting the mode-function
representation of $R_k$ gives
\begin{equation}\label{eq:YX-finite}
u_1\bigl(X_k-X_k^{\rm hom}\bigr)^*(t_1)
=\frac{2}{i}\int_{t_i}^{t_1}\dd t_2\,C_2\dot\alpha_2
\left[u_1^2u_2^*\dot u_2^*-|u_1|^2u_2\dot u_2^*\right].
\end{equation}
The real part of the first term on the right is $Y_k$ in
Eq.~\eqref{eq:Ymain}. For the second term the Wronskian gives
$C_2\mathrm{Im}(u_2\dot u_2^*)=1/4$, so
\begin{equation}\label{eq:Yidentity}
\begin{aligned}
Y_k={}&L_{01}+\frac H2(\partial_{\ln k}+3)L_{00}
+\frac{\alpha-\alpha_i}{2}L_{00}
-\mathrm{Re}\!\left[u_k(X_k^{\rm hom})^*\right].
\end{aligned}
\end{equation}
This reduces to Eq.~\eqref{eq:Ymain-identity} for the Bunch--Davies
initial scaling phase of section~\ref{sec:free}. A state
with a nonzero homogeneous dilation response carries an additional
term in the fish history. Similarly, nonzero $\alpha_i$ or
$\dot\alpha_i$ requires retaining the lower endpoints in the time
integrations by parts. These terms must be combined with the
corresponding transformation of the initial density matrix;
they are not removed by the bulk equation of motion.

\subsection{Moving upper limits and the complete later vertex}
The moving-upper-limit terms in Eq.~\eqref{eq:Ymain-leibniz}
follow from the equal-time identities in
Eq.~\eqref{eq:main-equal-time}: $h_k|_{\rm eq}=L_{00}/2$
produces the subtraction $-\dot\alpha L_{00}/2$, while
$\partial_{t_1}h_k|_{\rm eq}=0$ prevents a further endpoint
term in the second line.
Equation~\eqref{eq:Ymain-identity} and
$\dot L_{00}=2L_{01}$ imply
\begin{equation}
\dot Y_k=\dot L_{01}+H(\mathscr M_k+3)L_{01}
+\alpha L_{01}+\tfrac12\dot\alpha L_{00}
+\order(\dot H).
\end{equation}
Using Eq.~\eqref{eq:L-eom-main} gives the reduction to
$\mathcal B_k$ in Eq.~\eqref{eq:Y-to-B} and hence the complete
fish bulk in Eq.~\eqref{eq:fishbulkbody}.

Expanding its first term as
$-R_0\partial_t(\alpha C\mathcal B_k)
+R_0C\dot\alpha\mathcal B_k$
cancels the quartic $-R_0C\dot\alpha\mathcal D_k$,
leaving $R_0C\dot\alpha H\partial_{\ln k}L_{01}$.
The remaining total derivative contributes, after integration by
parts, $+\alpha\mathcal B_k/2$; the second term of
Eq.~\eqref{eq:fishbulkbody} contributes $-\alpha\mathcal B_k$.
Their sum is $-\alpha\mathcal B_k/2$.  Keeping the quartic mixed
contraction from Eq.~\eqref{eq:mixedsoft} then gives
Eq.~\eqref{eq:fishquartic}.

\Needspace{9\baselineskip}
\section{Non-1PI tadpoles and symmetry-related counterterms}\label{app:tad}
\subsection{Differentiated bridge and equal-time contact}
The factors in Eq.~\eqref{eq:tadfactors} reorganize the direct
non-1PI contractions into a one-loop zero-mode insertion
$\pi_{\rm 1L}^{\rm tad}$ and a differentiated-bridge insertion
$(\dot\pi)_{\rm 1L}^{\rm tad}$. The latter differentiates the bridge
at its later end with the integration domain held fixed.
With $\mathcal D_k$ from Eq.~\eqref{eq:SDdefs}, their bare forms are
\begin{align}\label{eq:tadzero}
\pi_{\rm 1L}^{\rm tad}(t_1)
&=\int_{t_i}^{t_1}\dd t_2\intk
\left[-C_2\alpha_2\mathcal D_kR_0(t_1,t_2)
-2C_2\alpha_2L_{01}\partial_{t_2}R_0(t_1,t_2)\right],\\
(\dot\pi)_{\rm 1L}^{\rm tad}(t_1)
&=\int_{t_i}^{t_1}\dd t_2\intk
\left[-C_2\alpha_2\mathcal D_k\partial_{t_1}R_0(t_1,t_2)
-2C_2\alpha_2L_{01}
\partial_{t_1}\partial_{t_2}R_0(t_1,t_2)\right].
\end{align}
The $L_{ab}$ in these integrals are evaluated at $(k,t_2)$.
To obtain the source of the one-point function, integrate the
second term of the first line by parts in $t_2$.  Its upper
endpoint vanishes because $R_0(t_1,t_1)=0$, while adiabatic
switching removes the lower endpoint.  The remaining integrand is
\begin{equation}
-C\alpha\mathcal D_kR_0
+2R_0\partial_t(C\alpha L_{01})
=CR_0\left[\alpha\mathcal D_k+2\dot\alpha L_{01}\right],
\end{equation}
where $\partial_t(CL_{01})=C\mathcal D_k$ was used in the
last equality.  This gives
\begin{equation}\label{eq:tadsource}
\pi_{\rm 1L}^{\rm tad}(t_1)
=\int_{t_i}^{t_1}\dd t_2\,C_2R_0(t_1,t_2)\Sigma_\pi(t_2).
\end{equation}
By contrast, differentiating the \emph{whole} first line of
Eq.~\eqref{eq:tadzero} with respect to its observation time
$t_1$ produces both the differentiated-bridge integral and an
upper-limit term. At equal time $R_0(t_1,t_1)=0$, while the
$\partial_{t_2}R_0$ term gives
\begin{equation}
\left[-2C_2\alpha_2L_{01}(t_2)
\partial_{t_2}R_0(t_1,t_2)\right]_{t_2=t_1^-}
=-\alpha_1L_{01}(t_1).
\end{equation}
Thus $\partial_{t_1}\pi_{\rm 1L}^{\rm tad}
=(\dot\pi)_{\rm 1L}^{\rm tad}-\alpha J$, or
\begin{equation}\label{eq:tadcontact}
(\dot\pi)_{\rm 1L}^{\rm tad}
=\partial_t\pi_{\rm 1L}^{\rm tad}+\alpha J.
\end{equation}
This $\alpha J$ is an equal-time canonical contact.  In the strict
soft limit the two-point tadpole is
\begin{equation}
\delta P_{\rm tad/1PR}^{\rm IR}
=-2\Ptree\int_{t_i}^{t}\dd t_1\,
\alpha_1(\dot\pi)_{\rm 1L}^{\rm tad}(t_1).
\end{equation}
Indeed, the finite-$p$ two-point expression obtained from
Eq.~\eqref{eq:tadkernels} is
\begin{equation}
\begin{aligned}
\delta P_{\rm tad/1PR}
=\int_{t_i}^{t}\dd t_1\,2\mathrm{Im}\big\{
&[2g_{T1}E_1^2+2g_{S1}p^2E_0^2]
\pi_{\rm 1L}^{\rm tad}(t_1)\\
&+4g_{T1}E_0E_1(\dot\pi)_{\rm 1L}^{\rm tad}(t_1)
\big\}.
\end{aligned}
\end{equation}
The first bracket contains either $p^2$ or
$\partial_{t_1}F_p$, so it is subleading.  The second retains
$F_p\partial_{t_1}R_0$ and gives the displayed soft result
after $g_T=-C\alpha$ and $C\partial_{t_1}R_0=1/2$ are used.

\subsection{One-point renormalization and its two-point insertions}
To fix the counterterms, take a single potential in the spatially
flat variable $\delta\phi$,
\begin{equation}
S_{\rm ct}^{\delta\phi}=\int\dd t\,\dd^3x\,a^3
\left[A_1(t)\delta\phi+A_2(t)(\delta\phi)^2\right],
\qquad \dot A_1=2\Pi A_2,\quad \Pi=\dot{\bar\phi}.
\end{equation}
Using $\delta\phi=\Pi\pi+\dot\Pi\pi^2/2+\cdots$ and
$A=\Pi^2/2$, the Goldstone-basis coefficients in
$S_{\rm ct}^{\pi}=\int a^3(b_1\pi+b_2\pi^2)$ obey
\begin{equation}\label{eq:bconstraint}
b_1=\Pi A_1,\qquad
b_2=\tfrac12\dot\Pi A_1+\Pi^2A_2
=\tfrac12\dot b_1.
\end{equation}
The linear insertion contributes
$\pi_{\rm ct}=-\int_{t_i}^t\dd t_1\,a_1^3b_1R_0(t,t_1)$.
The condition $\pi_{\rm 1L}^{\rm tad}+\pi_{\rm ct}=0$ for all
observation times implements a vanishing one-point renormalization
condition~\cite{KristianoYokoyama2025Background}. It
compares the two sources under the same retarded
kernel and fixes
\begin{equation}
a^3b_1=C\Sigma_\pi,\qquad b_1=A\Sigma_\pi,\qquad
b_2=\tfrac A2(\dot\Sigma_\pi+\alpha\Sigma_\pi).
\end{equation}
The counterterm is algebraic in $\pi$, giving
$(\dot\pi)_{\rm ct}=\partial_t\pi_{\rm ct}$; thus
Eq.~\eqref{eq:tadcontact} leaves
$(\dot\pi)_{\rm 1L}^{\rm tad}+(\dot\pi)_{\rm ct}=\alpha J$.
The corresponding soft two-point contribution is
$-2\Ptree\int\dd t_1\,\alpha_1^2J_1$.
The quadratic counterterm contributes
\begin{equation}
\delta P_{b_2}^{\rm IR}
=-2\Ptree\int_{t_i}^{t}\dd t_1\,
C_1R_0(t,t_1)(\dot\Sigma_{\pi,1}+\alpha_1\Sigma_{\pi,1}),
\end{equation}
completing Eq.~\eqref{eq:combined-loop-ct}.

\subsection{Reduction of the loop source}\label{app:source-reduction}
We give the steps from Eq.~\eqref{eq:combined-loop-ct} to
Eq.~\eqref{eq:result}. Throughout this reduction, a time derivative
of a loop integral means the integral of the time derivative of its
integrand at fixed $k$. No value is assigned to a momentum total
derivative.

The mode equation~\eqref{eq:mode} implies the equal-time relation
$\dot L_{01}=\mathcal D_k-(3H+\alpha)L_{01}$, with
$\mathcal D_k$ defined in Eq.~\eqref{eq:SDdefs}. Integrating at
fixed momentum gives
\begin{equation}\label{eq:loop-D-integral}
\intk\mathcal D_k=\dot J+(3H+\alpha)J.
\end{equation}
The definition $\mathcal B_k=\mathcal D_k+
H\partial_{\ln k}L_{01}$ then yields
\begin{equation}\label{eq:loop-source-identities}
\begin{aligned}
B&=\dot J+(3H+\alpha)J+HM
=\dot J+\alpha J+HK_{\partial k},\\
\Sigma_\pi&=\alpha[\dot J+(3H+\alpha)J]
+2\dot\alpha J,\\
\alpha B+2\dot\alpha J&=\Sigma_\pi+\alpha HM.
\end{aligned}
\end{equation}
The last line follows by subtracting the second line from
$\alpha$ times the first; it is the combination needed below.

The first two terms in Eq.~\eqref{eq:combined-loop-ct} do not
multiply $R_0$. We put them under the same retarded kernel by
integrating by parts in $t_1$. Since
$C_1\partial_{t_1}R_0(t,t_1)=1/2$,
\begin{equation}\label{eq:local-to-retarded}
\int_{t_i}^t\dd t_1\left(-\frac{\alpha_1}{2}B_1
-\dot\alpha_1J_1\right)
=\int_{t_i}^t\dd t_1\,R_0(t,t_1)
\partial_{t_1}\left(C_1\alpha_1B_1
+2C_1\dot\alpha_1J_1\right).
\end{equation}
The upper time endpoint vanishes by $R_0(t,t)=0$. At the initial
endpoint the adiabatic switching used in section~\ref{sec:free}
gives $\alpha_i=\dot\alpha_i=0$. Thus the source multiplying
$R_0$ is
\begin{equation}\label{eq:loop-source-first}
\mathcal W=\partial_t(C\alpha B+2C\dot\alpha J)
+C\dot\alpha HM-C(\dot\Sigma_\pi+\alpha\Sigma_\pi).
\end{equation}

Substitute the last identity in Eq.~\eqref{eq:loop-source-identities}
into the first term. Using $\dot C=(3H+\alpha)C$, the terms
proportional to $\dot\Sigma_\pi$ cancel without expanding $B$:
\begin{align}\label{eq:source-reduction}
\mathcal W
&=\partial_t\!\left[C(\Sigma_\pi+\alpha HM)\right]
+C\dot\alpha HM-C(\dot\Sigma_\pi+\alpha\Sigma_\pi)
\nonumber\\
&=3CH\Sigma_\pi+\partial_t(C\alpha HM)+C\dot\alpha HM.
\end{align}
We now insert $M=K_{\partial k}-3J$. This separates the
momentum total derivative from terms involving $J$:
\begin{equation}\label{eq:source-J-split}
\mathcal W=\partial_t(C\alpha HK_{\partial k})
+CH\dot\alpha K_{\partial k}
+3\left[CH\Sigma_\pi-\partial_t(C\alpha HJ)
-CH\dot\alpha J\right].
\end{equation}
The two expressions compared in the square bracket follow from
Eq.~\eqref{eq:loop-source-identities}:
\begin{equation}\label{eq:local-source-cancel}
\begin{aligned}
CH\Sigma_\pi
&=CH\left[\alpha\dot J+\alpha(3H+\alpha)J
+2\dot\alpha J\right],\\
\partial_t(C\alpha HJ)+CH\dot\alpha J
&=CH\left[\alpha\dot J+\alpha(3H+\alpha)J
+2\dot\alpha J\right]+C\alpha\dot HJ.
\end{aligned}
\end{equation}
Thus the $\dot J$, $(3H+\alpha)J$, and $\dot\alpha J$ terms
cancel pairwise, leaving the algebraic remainder
\begin{equation}\label{eq:source-exact-remainder}
\mathcal W=\partial_t(C\alpha HK_{\partial k})
+CH\dot\alpha K_{\partial k}-3C\alpha\dot HJ.
\end{equation}
The final term has an additional explicit factor of $\dot H$ and
is beyond the accuracy set in section~\ref{sec:eft}; it is not
the complete correction at that order, since the mode identity
and history reduction already omit terms proportional to $\dot H$.
Dropping this remainder gives
the result quoted in Eq.~\eqref{eq:result} without selecting a
regularization prescription for the momentum integral.

\bibliographystyle{jhep-template/JHEP}
\bibliography{dilation_one_loop_references}

@article{KristianoYokoyama2024Transitions,
  author = {Kristiano, J. and Yokoyama, J.},
  title = {{Comparing sharp and smooth transitions of the second slow-roll parameter in single-field inflation}},
  journal = {JCAP},
  volume = {10},
  pages = {036},
  year = {2024},
  eprint = {2405.12145},
  archivePrefix = {arXiv},
}

@article{BarnertEtAl2026,
  author = {Barnert, G. and Iacconi, L. and Assadullahi, H. and Koyama, K. and Wands, D.},
  title = {{UV artefacts in ultra-slow-roll models of inflation}},
  year = {2026},
  eprint = {2606.28005},
  archivePrefix = {arXiv},
}

@article{MotohashiTada2023,
  author = {Motohashi, H. and Tada, Y.},
  title = {{Squeezed bispectrum and one-loop corrections in transient constant-roll inflation}},
  journal = {JCAP},
  volume = {08},
  pages = {069},
  year = {2023},
  eprint = {2303.16035},
  archivePrefix = {arXiv},
}

@article{FirouzjahiRiotto2024,
  author = {Firouzjahi, H. and Riotto, A.},
  title = {{Primordial Black Holes and loops in single-field inflation}},
  journal = {JCAP},
  volume = {02},
  pages = {021},
  year = {2024},
  eprint = {2304.07801},
  archivePrefix = {arXiv},
}

@article{BallesterosEgea2024,
  author = {Ballesteros, G. and Gamb\'in Egea, J.},
  title = {{One-loop power spectrum in ultra slow-roll inflation and implications for primordial black hole dark matter}},
  journal = {JCAP},
  volume = {07},
  pages = {052},
  year = {2024},
  eprint = {2404.07196},
  archivePrefix = {arXiv},
}

@article{KristianoYokoyama2024,
  author = {Kristiano, J. and Yokoyama, J.},
  title = {{Constraining Primordial Black Hole Formation from Single-Field Inflation}},
  journal = {Phys. Rev. Lett.},
  volume = {132},
  pages = {221003},
  year = {2024},
  eprint = {2211.03395},
  archivePrefix = {arXiv},
}

@article{KristianoYokoyama2024Bispectrum,
  author = {Kristiano, J. and Yokoyama, J.},
  title = {{Note on the bispectrum and one-loop corrections in single-field inflation with primordial black hole formation}},
  journal = {Phys. Rev. D},
  volume = {109},
  pages = {103541},
  year = {2024},
  eprint = {2303.00341},
  archivePrefix = {arXiv},
}

@article{Firouzjahi2023,
  author = {Firouzjahi, H.},
  title = {{One-loop corrections in power spectrum in single field inflation}},
  journal = {JCAP},
  volume = {10},
  pages = {006},
  year = {2023},
  eprint = {2303.12025},
  archivePrefix = {arXiv},
}

@article{FrancioliniEtAl2024,
  author = {Franciolini, G. and Iovino, Jr., A. and Taoso, M. and Urbano, A.},
  title = {{Perturbativity in the presence of ultraslow-roll dynamics}},
  journal = {Phys. Rev. D},
  volume = {109},
  pages = {123550},
  year = {2024},
  eprint = {2305.03491},
  archivePrefix = {arXiv},
}

@article{Firouzjahi2024Revisiting,
  author = {Firouzjahi, H.},
  title = {{Revisiting loop corrections in single field ultraslow-roll inflation}},
  journal = {Phys. Rev. D},
  volume = {109},
  pages = {043514},
  year = {2024},
  eprint = {2311.04080},
  archivePrefix = {arXiv},
}

@article{ChengLeeNg2024,
  author = {Cheng, S.-L. and Lee, D.-S. and Ng, K.-W.},
  title = {{Primordial perturbations from ultra-slow-roll single-field inflation with quantum loop effects}},
  journal = {JCAP},
  volume = {03},
  pages = {008},
  year = {2024},
  eprint = {2305.16810},
  archivePrefix = {arXiv},
}

@article{MaityEtAl2024,
  author = {Maity, S. and Ragavendra, H. V. and Sethi, S. K. and Sriramkumar, L.},
  title = {{Loop contributions to the scalar power spectrum due to quartic order action in ultra slow roll inflation}},
  journal = {JCAP},
  volume = {05},
  pages = {046},
  year = {2024},
  eprint = {2307.13636},
  archivePrefix = {arXiv},
}

@article{Riotto2023,
  author = {Riotto, A.},
  title = {{The Primordial Black Hole Formation from Single-Field Inflation is Not Ruled Out}},
  year = {2023},
  eprint = {2301.00599},
  archivePrefix = {arXiv},
}

@article{TadaEtAl2024,
  author = {Tada, Y. and Terada, T. and Tokuda, J.},
  title = {{Cancellation of quantum corrections on the soft curvature perturbations}},
  journal = {JHEP},
  volume = {01},
  pages = {105},
  year = {2024},
  eprint = {2308.04732},
  archivePrefix = {arXiv},
}

@article{Inomata2024,
  author = {Inomata, K.},
  title = {{Superhorizon Curvature Perturbations Are Protected against One-Loop Corrections}},
  journal = {Phys. Rev. Lett.},
  volume = {133},
  pages = {141001},
  year = {2024},
  eprint = {2403.04682},
  archivePrefix = {arXiv},
}

@article{KawaguchiEtAl2024,
  author = {Kawaguchi, R. and Tsujikawa, S. and Yamada, Y.},
  title = {{Proving the absence of large one-loop corrections to the power spectrum of curvature perturbations in transient ultra-slow-roll inflation within the path-integral approach}},
  journal = {JHEP},
  volume = {12},
  pages = {095},
  year = {2024},
  eprint = {2407.19742},
  archivePrefix = {arXiv},
}

@article{Cheung2008,
  author = {Cheung, C. and Creminelli, P. and Fitzpatrick, A. L. and Kaplan, J. and Senatore, L.},
  title = {{The Effective Field Theory of Inflation}},
  journal = {JHEP},
  volume = {03},
  pages = {014},
  year = {2008},
  eprint = {0709.0293},
  archivePrefix = {arXiv},
}

@article{ChoudhuryPandaSami2023,
  author = {Choudhury, S. and Panda, S. and Sami, M.},
  title = {{PBH formation in EFT of single field inflation with sharp transition}},
  journal = {Phys. Lett. B},
  volume = {845},
  pages = {138123},
  year = {2023},
  eprint = {2302.05655},
  archivePrefix = {arXiv},
}

@article{FirouzjahiNikbakht2025,
  author = {Firouzjahi, H. and Nikbakht, B.},
  title = {{Nonperturbative Hamiltonian and higher loop corrections in ultraslow-roll inflation}},
  journal = {Phys. Rev. D},
  volume = {113},
  pages = {L061301},
  year = {2026},
  eprint = {2502.09481},
  archivePrefix = {arXiv},
}

@article{FirouzjahiNikbakht2026AllOrders,
  author = {Firouzjahi, H. and Nikbakht, B.},
  title = {{Hamiltonians to all orders in perturbation theory and higher-loop corrections in single-field inflation with PBH formation}},
  journal = {Phys. Rev. D},
  volume = {113},
  pages = {063510},
  year = {2026},
  eprint = {2502.10287},
  archivePrefix = {arXiv},
}

@article{BragliaPinol2025,
  author = {Braglia, M. and Pinol, L.},
  title = {{One-loop renormalization of the effective field theory of inflationary fluctuations from gravitational interactions}},
  journal = {Phys. Rev. D},
  volume = {113},
  pages = {063513},
  year = {2026},
  eprint = {2504.07926},
  archivePrefix = {arXiv},
}

@article{BragliaPinol2026Freezing,
  author = {Braglia, M. and Pinol, L.},
  title = {{Freezing of the renormalized one-loop primordial scalar power spectrum}},
  journal = {Phys. Rev. D},
  volume = {113},
  pages = {L061302},
  year = {2026},
  eprint = {2504.13136},
  archivePrefix = {arXiv},
}

@article{BragliaCespedesPinol2026,
  author = {Braglia, M. and C\'espedes, S. and Pinol, L.},
  title = {{Scale-dependent loop corrections to the inflationary power spectrum}},
  journal = {JHEP},
  volume = {07},
  pages = {250},
  year = {2026},
  eprint = {2603.12216},
  archivePrefix = {arXiv},
}

@article{Maldacena2003,
  author = {Maldacena, J. M.},
  title = {{Non-Gaussian features of primordial fluctuations in single field inflationary models}},
  journal = {JHEP},
  volume = {05},
  pages = {013},
  year = {2003},
  eprint = {astro-ph/0210603},
  archivePrefix = {arXiv},
}

@article{CreminelliZaldarriaga2004,
  author = {Creminelli, P. and Zaldarriaga, M.},
  title = {{Single field consistency relation for the 3-point function}},
  journal = {JCAP},
  volume = {10},
  pages = {006},
  year = {2004},
  eprint = {astro-ph/0407059},
  archivePrefix = {arXiv},
}

@article{HinterbichlerEtAl2012,
  author = {Hinterbichler, K. and Hui, L. and Khoury, J.},
  title = {{Conformal Symmetries of Adiabatic Modes in Cosmology}},
  journal = {JCAP},
  volume = {08},
  pages = {017},
  year = {2012},
  eprint = {1203.6351},
  archivePrefix = {arXiv},
}

@article{GoldbergerHuiNicolis2013,
  author = {Goldberger, W. D. and Hui, L. and Nicolis, A.},
  title = {{One-particle-irreducible consistency relations for cosmological perturbations}},
  journal = {Phys. Rev. D},
  volume = {87},
  pages = {103520},
  year = {2013},
  eprint = {1303.1193},
  archivePrefix = {arXiv},
}

@article{HinterbichlerEtAl2014,
  author = {Hinterbichler, K. and Hui, L. and Khoury, J.},
  title = {{An Infinite Set of Ward Identities for Adiabatic Modes in Cosmology}},
  journal = {JCAP},
  volume = {01},
  pages = {039},
  year = {2014},
  eprint = {1304.5527},
  archivePrefix = {arXiv},
}

@article{PimentelEtAl2012,
  author = {Pimentel, G. L. and Senatore, L. and Zaldarriaga, M.},
  title = {{On Loops in Inflation III: Time Independence of zeta in Single Clock Inflation}},
  journal = {JHEP},
  volume = {07},
  pages = {166},
  year = {2012},
  eprint = {1203.6651},
  archivePrefix = {arXiv},
}

@article{AssassiEtAl2013,
  author = {Assassi, V. and Baumann, D. and Green, D.},
  title = {{Symmetries and Loops in Inflation}},
  journal = {JHEP},
  volume = {02},
  pages = {151},
  year = {2013},
  eprint = {1210.7792},
  archivePrefix = {arXiv},
}

@article{SenatoreZaldarriaga2013AllLoops,
  author = {Senatore, L. and Zaldarriaga, M.},
  title = {{The constancy of $\zeta$ in single-clock Inflation at all loops}},
  journal = {JHEP},
  volume = {09},
  pages = {148},
  year = {2013},
  eprint = {1210.6048},
  archivePrefix = {arXiv},
}

@article{Fumagalli2025PartI,
  author = {Fumagalli, J.},
  title = {{Absence of one-loop effects on large scales from small scales in non-slow-roll dynamics}},
  journal = {JHEP},
  volume = {05},
  pages = {162},
  year = {2025},
  eprint = {2305.19263},
  archivePrefix = {arXiv},
}

@article{Fumagalli2025,
  author = {Fumagalli, J.},
  title = {{Absence of one-loop effects on large scales from small scales in non-slow-roll dynamics. Part 2. Quartic interactions and consistency relations}},
  journal = {JHEP},
  volume = {01},
  pages = {108},
  year = {2025},
  eprint = {2408.08296},
  archivePrefix = {arXiv},
}

@article{ArtigasEtAl2026,
  author = {Artigas, D. and Pi, S. and Tanaka, T. and Urakawa, Y.},
  title = {{How Dilatation Invariance Suppresses Loop Corrections to Curvature Perturbations on CMB Scales}},
  year = {2026},
  eprint = {2608.03725},
  archivePrefix = {arXiv},
}

@article{FangLyuChenGuo2026,
  author = {Fang, C.-J. and Lyu, Z.-H. and Chen, C. and Guo, Z.-K.},
  title = {{Ward identity constraints on loop corrections in non-attractor inflation}},
  journal = {Phys. Rev. D},
  volume = {114},
  pages = {L021306},
  year = {2026},
  eprint = {2507.00077},
  archivePrefix = {arXiv},
}

@article{FangGuo2026,
  author = {Fang, C.-J. and Guo, Z.-K.},
  title = {{Fixing the Renormalization of Inflationary Loops via Ward Identities}},
  journal = {JHEP},
  volume = {10},
  pages = {121},
  year = {2026},
  eprint = {2605.13325},
  archivePrefix = {arXiv},
}

@article{Inomata2025Backreaction,
  author = {Inomata, K.},
  title = {{Conservation of superhorizon curvature perturbations at one loop: Backreaction in the in-in formalism and renormalization}},
  journal = {Phys. Rev. D},
  volume = {111},
  pages = {103504},
  year = {2025},
  eprint = {2502.08707},
  archivePrefix = {arXiv},
}

@article{FangEtAl2025Backreaction,
  author = {Fang, C.-J. and Lyu, Z.-H. and Chen, C. and Guo, Z.-K.},
  title = {{Incorporating backreaction in one-loop corrections in ultraslow-roll inflation}},
  journal = {Phys. Rev. D},
  volume = {112},
  pages = {023547},
  year = {2025},
  eprint = {2502.09555},
  archivePrefix = {arXiv},
}

@article{Inomata2025Counterterms,
  author = {Inomata, K.},
  title = {{Role of the counterterms in the conservation of superhorizon curvature perturbations at one loop}},
  journal = {Phys. Rev. D},
  volume = {111},
  pages = {123517},
  year = {2025},
  eprint = {2502.12112},
  archivePrefix = {arXiv},
}

@article{KristianoYokoyama2025Background,
  author = {Kristiano, J. and Yokoyama, J.},
  title = {{Inflationary background renormalization}},
  year = {2025},
  eprint = {2504.18514},
  archivePrefix = {arXiv},
}

@article{IacconiEtAl2024,
  author = {Iacconi, L. and Mulryne, D. and Seery, D.},
  title = {{Loop corrections in the separate universe picture}},
  journal = {JCAP},
  volume = {06},
  pages = {062},
  year = {2024},
  eprint = {2312.12424},
  archivePrefix = {arXiv},
}

@article{IacconiEtAl2026,
  author = {Iacconi, L. and Mulryne, D. and Seery, D.},
  title = {{Decoupling of large-scale, adiabatic inflationary perturbations from enhanced small-scale modes at one-loop}},
  journal = {JCAP},
  volume = {08},
  pages = {034},
  year = {2026},
  eprint = {2601.14229},
  archivePrefix = {arXiv},
}

@article{EmaEtAl2026,
  author = {Ema, Y. and Hong, M. and Jinno, R. and Mukaida, K.},
  title = {{Cancellation of loop corrections to soft scalar power spectrum}},
  year = {2026},
  eprint = {2603.01961},
  archivePrefix = {arXiv},
}

@article{TanakaUrakawa2026,
  author = {Tanaka, T. and Urakawa, Y.},
  title = {{Locality in effective field theory for inflationary soft modes}},
  year = {2026},
  eprint = {2605.19331},
  archivePrefix = {arXiv},
}

@article{BragliaPinol2024,
  author = {Braglia, M. and Pinol, L.},
  title = {{No time to derive: unraveling total time derivatives in in-in perturbation theory}},
  journal = {JHEP},
  volume = {08},
  pages = {068},
  year = {2024},
  eprint = {2403.14558},
  archivePrefix = {arXiv},
}

@article{ArrojaTanaka2011,
  author = {Arroja, F. and Tanaka, T.},
  title = {{A note on the role of the boundary terms for the non-Gaussianity in general k-inflation}},
  journal = {JCAP},
  volume = {05},
  pages = {005},
  year = {2011},
  eprint = {1103.1102},
  archivePrefix = {arXiv},
}

@article{WangEtAl2025FieldRedefinitions,
  author = {Wang, D. and Wang, X. and Wang, Y. and Yu, W.},
  title = {{Interact or Twist: Cosmological Correlators from Field Redefinitions Revisited}},
  year = {2025},
  eprint = {2508.12856},
  archivePrefix = {arXiv},
}

@article{MaEtAl2026Boundary,
  author = {Ma, Y. and Wang, D.-G. and Wang, X. and Wang, Y. and Yu, W.},
  title = {{On Cosmological Correlators with Boundary Contributions}},
  year = {2026},
  eprint = {2606.06282},
  archivePrefix = {arXiv},
}

@article{Creminelli:2024cge,
  author = {Creminelli, P. and Renaux-Petel, S. and Tambalo, G. and Yingcharoenrat, V.},
  title = {{Non-perturbative wavefunction of the universe in inflation with (resonant) features}},
  journal = {JHEP},
  volume = {03},
  pages = {010},
  year = {2024},
  doi = {10.1007/JHEP03(2024)010},
  eprint = {2401.10212},
  archivePrefix = {arXiv},
  primaryClass = {hep-th},
}

@article{BehbahaniEtAl2012,
  author = {Behbahani, S. R. and Dymarsky, A. and Mirbabayi, M. and Senatore, L.},
  title = {{(Small) Resonant non-Gaussianities: Signatures of a Discrete Shift Symmetry in the Effective Field Theory of Inflation}},
  journal = {JCAP},
  volume = {12},
  pages = {036},
  year = {2012},
  eprint = {1111.3373},
  archivePrefix = {arXiv},
}

@article{Weinberg2005,
  author = {Weinberg, S.},
  title = {{Quantum contributions to cosmological correlations}},
  journal = {Phys. Rev. D},
  volume = {72},
  pages = {043514},
  year = {2005},
  eprint = {hep-th/0506236},
  archivePrefix = {arXiv},
}

@article{Weinberg2006,
  author = {Weinberg, S.},
  title = {{Quantum contributions to cosmological correlations. II. Can these corrections become large?}},
  journal = {Phys. Rev. D},
  volume = {74},
  pages = {023508},
  year = {2006},
  eprint = {hep-th/0605244},
  archivePrefix = {arXiv},
}

@article{SenatoreZaldarriaga2010,
  author = {Senatore, L. and Zaldarriaga, M.},
  title = {{On Loops in Inflation}},
  journal = {JHEP},
  volume = {12},
  pages = {008},
  year = {2010},
  eprint = {0912.2734},
  archivePrefix = {arXiv},
}

@article{HuiEtAl2019,
  author = {Hui, L. and Joyce, A. and Wong, S. S. C.},
  title = {{Inflationary soft theorems revisited: A generalized consistency relation}},
  journal = {JCAP},
  volume = {02},
  pages = {060},
  year = {2019},
  eprint = {1811.05951},
  archivePrefix = {arXiv},
}

@article{SheikhahmadiNassiriRad2024,
  author = {Sheikhahmadi, H. and Nassiri-Rad, A.},
  title = {{RG-Flow Renormalized One-Loop Corrections to the Power Spectrum in USR Inflation}},
  year = {2024},
  eprint = {2411.18525},
  archivePrefix = {arXiv},
}

@article{BallesterosEtAl2025Finite,
  author = {Ballesteros, G. and Gamb\'in Egea, J. and Riccardi, F.},
  title = {{Finite parts of inflationary loops}},
  journal = {JHEP},
  volume = {06},
  pages = {098},
  year = {2025},
  eprint = {2411.19674},
  archivePrefix = {arXiv},
}

@article{BallesterosEtAl2026Finite,
  author = {Ballesteros, G. and Gamb\'in Egea, J. and Riccardi, F.},
  title = {{Finite parts of inflationary loops II: A streamlined UV in-in algorithm and distinguishable signatures}},
  journal = {JHEP},
  volume = {07},
  pages = {059},
  year = {2026},
  eprint = {2512.20467},
  archivePrefix = {arXiv},
}

@article{Inomata2026AllScales,
  author = {Inomata, K.},
  title = {{Cancellation of one-loop time dependence in superhorizon curvature perturbations from all scales}},
  year = {2026},
  eprint = {2606.28247},
  archivePrefix = {arXiv},
}

\end{document}